\documentstyle[10pt,fullpage,psfig,bbbold,twocolum]{article}

\def\slantfrac#1#2{\hbox{$\,^{#1}\!/_{#2}$}}

\begin{document}

\title{On possible observational evidences of spiral shocks in
accretion disks of CVs}

\author{D.V.Bisikalo$^1$, A.A.Boyarchuk$^1$,
A.A.Kilpio$^1$, O.A.Kuznetsov$^2$\\[0.5cm]
$^1$ {\it Institute of Astronomy of the Russian Acad. of Sci.,
Moscow}\\
{\sf bisikalo@inasan.rssi.ru; aboyar@inasan.rssi.ru;
skilpio@inasan.rssi.ru}\\[0.3cm]
$^2$ {\it Keldysh Institute of Applied Mathematics, Moscow}\\
{\sf kuznecov@spp.keldysh.ru}\\[0.3cm]}
\date{}
\maketitle

\begin{abstract}

The results of three-dimensional numerical simulations of mass
transfer in binaries with spiral shocks are presented. It is
shown that the mass transfer rate variation disturbs the
equilibrium state of accretion disk and results in formation of
the blob. This dense formation moves on the disk with variable
velocity. The results of our simulations show that the blob
lives long enough and retains its main characteristics for the
time of order of tens orbital periods. Analysis of the results
shows that the action of dissipation is negligible on these
timescales and the blob smearing out due to differential
rotation of the disk is stopped by its interaction with spiral
shocks.

Light curves of some CVs show periodic or quasi-periodic
photometric modulations with typical width $\sim0^{\rm
P}\!\!.1\div0^{\rm P}\!\!.2$ (i.e. with period
$\sim0.1\div0.2P_{orb}$). We suggest to consider the formation of
the blob in binaries with spiral shocks as a possible reason for
these modulations. The efficiency of this mechanism is confirmed
by good qualitative (and, in part, quantitative) agreement
between the results of our simulations and observations of
binaries IP Peg and EX Dra.

\end{abstract}

\section{Introduction}

Analysis of light curves in CVs with high orbit inclination and
especially eclipsing binaries permits to get information on the
flow structure between the components of the system. The shape of
light curves in these binaries is very sophisticated and even
uneclipsed parts of light curve show significant brightness
oscillations (see, e.g., Hack \& La Dous
1993$^{\cite{ladous93}}$; Warner 1995$^{\cite{warner95}}$). A
part of these oscillations, the so called flickering has
aperiodic nature and is characterized by small magnitude and
short timescale. On the other hand, besides sporadic
short-period oscillations light curves in CVs demonstrate
periodic or quasi-periodic photometric modulations with typical
period $\sim0.1\div0.2P_{orb}$. Various models were suggested to
explain these modulations of light curves: from intrinsic disk
instabilities up to modulations of the rate of mass transfer
from the mass-donating star (Warner 1995$^{\cite{warner95}}$).
Based on the results of 3D gasdynamical simulations we concluded
in (Bisikalo et al. 2001$^{\cite{blob1}}$, hereinafter Paper I)
that if spiral shocks are presented in the accretion disk then
any disturbance of the disk would result in the appearance a new
dense formation in the disk -- a blob, the latter moving through
the disk with variable velocity. A significant density contrast
between the blob and the disk (up to 1.5 times) as well as the
variable velocity of the blob (but with constant period of
revolution $\sim0.1\div0.2P_{orb}$) permit us to consider the
appearance of this formation as a possible reason for observable
quasi-periodic oscillations on the light curves in CVs. If the
emissivity of the blob is larger than that for the accretion
disk (e.g., for the case of recombination radiation which is
proportional to the square of density) the oscillations on the
light curve can be caused by the occultation the blob by
optically thick disk. And if the blob emissivity is less than
that for the disk the oscillations can be caused by occultation
of the disk by optically thick blob.

The main aim of this work is more detailed (as compared to Paper
I) investigation of the blob formation as well as a
determination of the dependence of its features on the
parameters of the model. We also analyze observations of
eclipsed binaries with spiral shocks (IP Peg and EX Dra) in the
context of applicability of proposed mechanism for explanation
of oscillations on light curves in these binaries.

\section{Results of numerical\\ simulations in CVs with variable
mass transfer rate}

In Paper I we have presented the results of 3D gasdynamical
simulations of flow structure in semidetached binaries after the
mass transfer termination. The investigation of the residual
accretion disk shows that as early as at the time
$t=t_0+0.2P_{orb}$ (where $t_0$ is time when mass transfer has
been ceased) the flow structure is changed significantly. The
stream from $L_1$ doesn't dominate anymore, and the shape of
accretion disk changes from quasi-elliptical to circular. The
second arm of tidally induced spiral shock is formed while
earlier (before the termination of mass transfer) it was
suppressed by the stream from $L_1$. The results of numerical
simulations presented in Paper I show that a dense blob is
formed in the residual disk, the velocity of the blob motion
through the disk being variable. The blob is a dense formation
which in steady-state solution coincides with a post-shock zone
of the spiral shock. After mass transfer termination when the
nourishment of the disk from the stream is ceased and equilibria
of the disk is disturbed the blob comes off from the front of
spiral shock and begins to move.

Two factors can force the blob to smear out as it moves through
the accretion disk -- dissipation and differential rotation of
the disk. Dissipation smearing occurs in viscous timescale which
can be evaluated as tens --- hundreds orbital periods for
$\alpha\sim0.01\div0.1$ which is typical for CVs.\footnote{
$$
\tau_\nu=\frac{R_d^2}{\nu}=\frac{R_d^2}{\alpha cH}=
\frac{R_d V_K}{\alpha c^2}
\approx\frac{1}{\alpha\beta^2\Omega_K}\,,
$$
where $R_d$ -- the radius of accretion disk; $\nu$ -- kinematic
viscosity; $c$ -- sound speed; $\alpha$ -- Shakura-Sunyaev
parameter; $H$ -- height scale of the disk; $\beta=H/R_d=c/V_K$;
$\Omega_K$ and $V_K$ -- angular and linear rotational velocities
of Keplerian disk. Transform the last expression to obtain
$$
\frac{\tau_\nu}{P_{orb}}=\frac{\Omega_{orb}}{2\pi\alpha\beta^2\Omega_K}
=\frac{\sqrt{1+q}}{2\pi\alpha\beta^2}\left(\frac{R_d}{A}\right)^{3/2}\,,
$$
where $P_{orb}$ -- orbital period; $q$ -- mass ratio (the ratio
of donor-star mass to accretor mass); $A$ -- the separation of
binary. In our calculations the typical values of parameters
are: $\beta\sim0.1\div0.15$ and $R_d/A\sim0.3\div0.4$, so for
$\alpha\sim0.1$ we have $\tau_\nu\sim25P_{orb}$ while for
$\alpha\sim0.01$ we have $\tau_\nu\sim250P_{orb}$. In Paper I we
have obtained the lifetimes of accretion disk for various values
of $\alpha$ which give a little bit less but comparable values
of $\tau_\nu$.} We would like to stress that the viscosity
timescale of smearing out of the blob gives the lower estimation
of its lifetime, since this time is so large that the recurrent
mass transfer rate variation and the subsequent disturbance of
the disk is very probable. It means that the new disturbance of
the disk due to mass transfer rate variation can occur earlier
than the dissipative smearing out of the blob.

Smearing out (or mixing) of the blob due to differential
rotation of the disk is virtually more important since this
process occurs in dynamical timescale and one may suppose that
the blob disappears during few orbital periods. Nevertheless, 3D
gasdynamical calculations of Paper I show that existence of a
spiral shock prevents dynamical mixing of the blob. Passing
through the front of spiral shock the blob becomes denser and
more compact, this effect being presented up to the moment when
spiral shock disappears. The density contrast between the blob
and the disk begins to decrease after the disappearing of the
spiral shock as well.

We understand that the situation of complete termination of mass
transfer from donor-star is rather rare, so we should consider
the possibility of the blob formation when the variation of the
mass transfer rate is relatively small. In this paper we present
the results of investigation of the flow structure in
semidetached binary after the mass transfer becomes weaker twice
and 10 times.

Note, that in the case of complete termination of mass transfer
our calculations have been stopped after the disk vanishes,
while in case of the reduced rate of mass transfer we have
observed the flow structure went out to a new quasi-steady-state
solution. In general, to reach an ultimate steady-state solution
we should conduct our simulations over time interval comparable
to viscous timescale of the disk. Nevertheless, we have
restricted this time to few tens of orbital periods bearing in
mind the typical timescale of outburst activity in CVs resulting
in the formation of spiral shocks.

The full description of the 3D gasdynamical model can be found
in Paper I. Here we pay attention only to the main features of
the model. To describe the gas flow in the binary we used the 3D
system of Euler equations for Cartesian coordinate system. To
close the system of equations, we used the equation of state of
ideal gas with adiabatic index $\gamma\sim1$, that corresponds
to the case close to the isothermal one. Other dimensionless
parameters which determine the solution were adopted as
follows: mass ratio is $q=1$, the ratio of sound speed in $L_1$
to orbital velocity is $\epsilon=c(L_1)/A\Omega$.  The gas flow
was simulated over a parallelepipedon $[\slantfrac{1}{2}A\ldots
\slantfrac{3}{2}A]\times[-\slantfrac{1}{2}A\ldots
\slantfrac{1}{2}A]\times[0\ldots\slantfrac{1}{4}A]$ without a
sphere with a radius of $\slantfrac{1}{100}A$ representing the
accretor. The calculations were conducted on the grid
$61\times61\times17$, in terms of the $\alpha$-disk the
numerical viscosity approximately corresponded to
$\alpha\sim0.05$. To obtain the numerical solution of the system
of equations we used the Roe--Osher TVD scheme of a high
approximation order (Roe 1986$^{\cite{roe86}}$; Chakravarthy \&
Osher 1985$^{\cite{osher85}}$). We used quasi-steady-state
solution for constant rate of mass transfer as initial
condition. At the moment of time $t=t_0$ we decreased the
density of the injected matter twice (run \lq\lq A") and 10 times (run
\lq\lq B").

\begin{figure}[t]
\centerline{\hbox{\psfig{figure=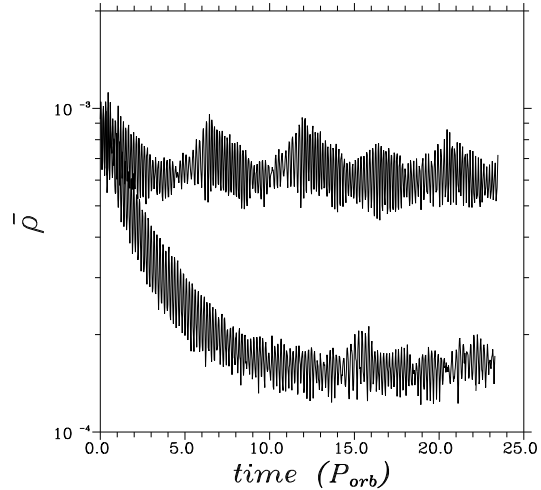,width=7cm}}}
\caption{\footnotesize Time variation of the mean density of
matter passing through a semi-plane $XZ$ ($Y=0$, $X>A$) slicing
the disk for run \lq\lq A" (upper curve) and \lq\lq B" (lower curve).}
\end{figure}

\begin{figure}[t]
\centerline{\hbox{\psfig{figure=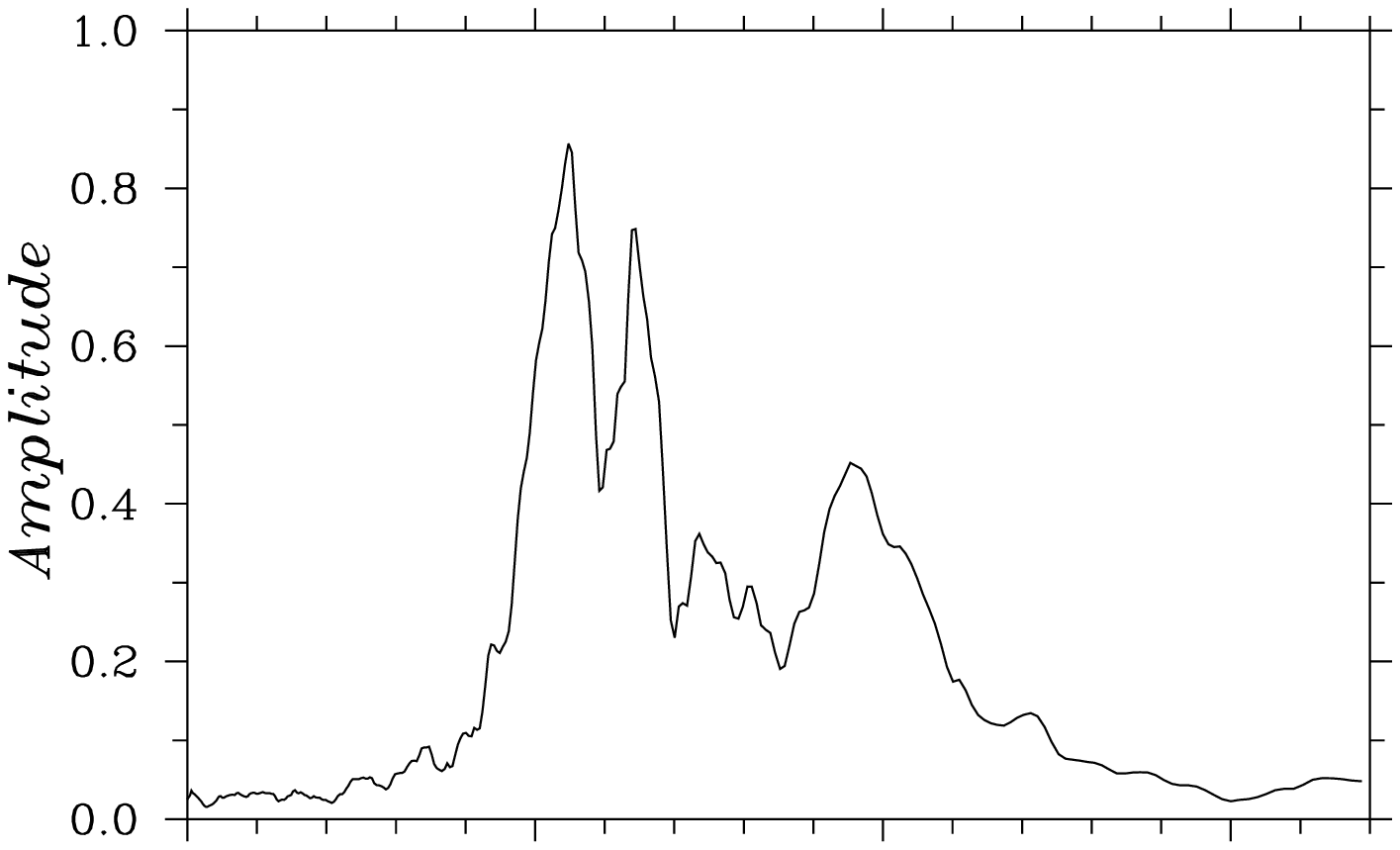,width=7cm}}}
\centerline{\hbox{\psfig{figure=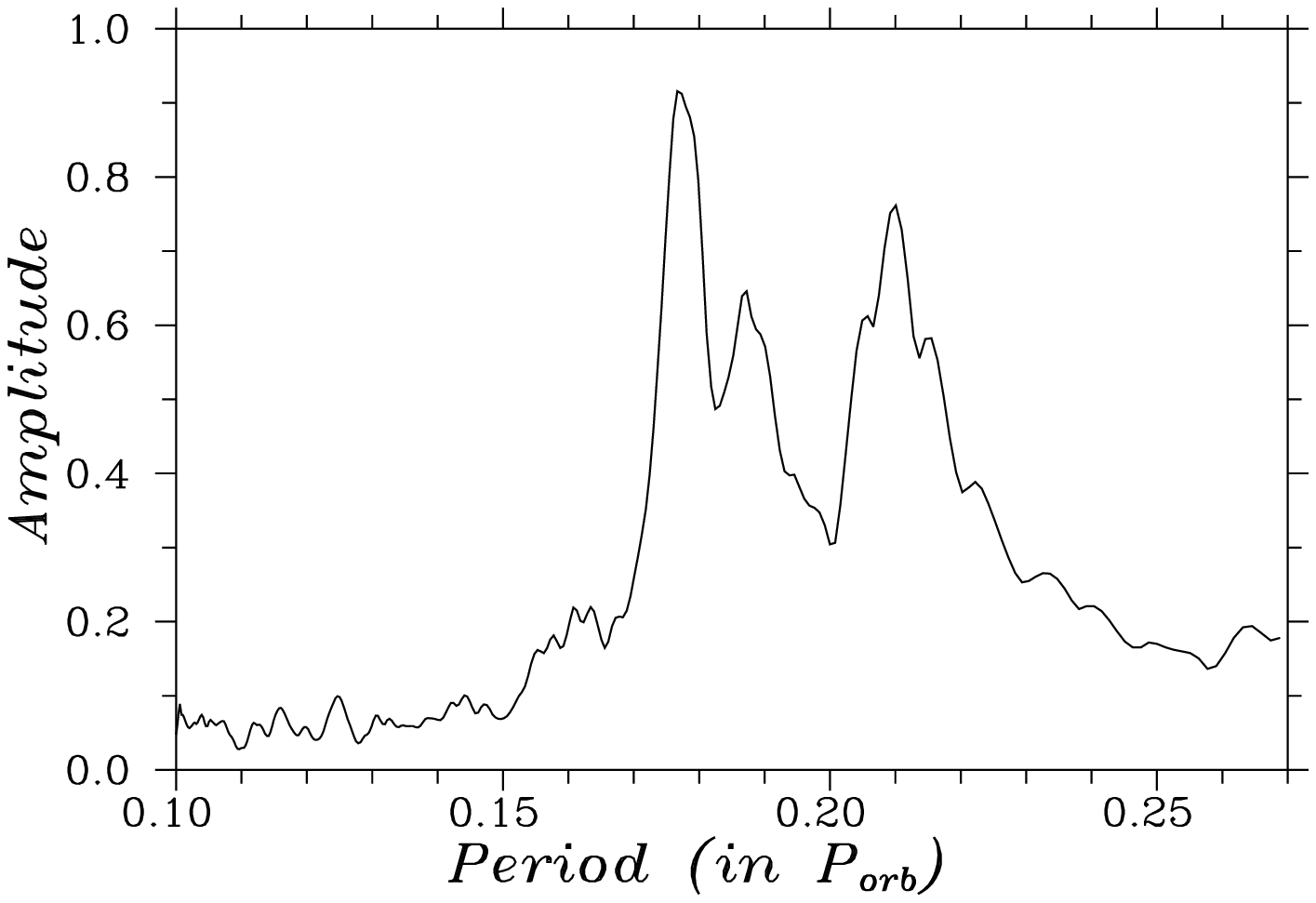,width=7cm}}}
\caption{\footnotesize Strength of Fourier harmonics versus
period for run \lq\lq A" (upper panel) and \lq\lq B" (lower
panel).}
\end{figure}

The Figure 1 presents time variation of the mean density of
matter passing through a semi-plane $XZ$ ($Y=0$, $X>A$) slicing
the disk. Along with a general density fall we can see here
periodic oscillations called forth by passing the blob. The
analysis of curves in Fig.~1 shows that after changing
(decreasing) the mass transfer rate a dense formation -- blob --
arises. The period of blob revolution is $\sim 0.15P_{orb}$ for
run \lq\lq A" and $\sim 0.18P_{orb}$ for run \lq\lq B". After recurrent
reaching of quasi-steady-state solution (i.e. for
$t>t_0+3P_{orb}$ for run \lq\lq A" and $t>t_0+13P_{orb}$ for run \lq\lq B")
the period of the blob revolution increases and becomes $\sim
0.16P_{orb}$ for run \lq\lq A" and $\sim 0.20P_{orb}$ for run \lq\lq B".
The new quasi-steady-state solution is characterized by
long-period oscillations (Fig.~1) with period $\sim 5.1P_{orb}$
for run \lq\lq A" and $\sim 2.5P_{orb}$ for run \lq\lq B". The Figure 2
presents the Fourier analysis of curves from Fig.~1. It is seen
that for run \lq\lq A" dominating harmonics are $0.155P_{orb}$ and
$0.165P_{orb}$ (Fig.~2a). The beating of these two gives
long-period oscillations with period $\sim 5.1P_{orb}$. For run
\lq\lq B" dominating harmonics are $0.18P_{orb}$ and $0.21P_{orb}$
(Fig.~2b), and their beating gives long-period oscillations with
the period $\sim 2.5P_{orb}$.

Let us consider how does the disk structure change after the
rate of mass transfer decreases. The Figure 3 presents the
results of run \lq\lq A" and shows the density distributions over the
equatorial plane for 6 moments of time $t_0+0.44P_{orb}$,
$0.47P_{orb}$, $0.49P_{orb}$, $0.52P_{orb}$, $0.54P_{orb}$,
$0.57P_{orb}$, covering the full period of the blob revolution.
The maximal density in Fig.~3 corresponds to
$\rho\simeq0.035\rho(L_1)$. These results are relevant to the
transient stage when dense stream from $L_1$ vanishes at the
moment of time $t=t_0$ and the flow structure is determined by
the relatively dense disk while the matter of rarefied stream
(with density twice less than the initial density) plays a minor
role. This stage begins after the last portions of matter of
the old stream come to the disk, i.e. at $t=t_0+0.2P_{orb}$, and
finishes when a new quasi-steady-state solution will be reached.
The main features of this stage are: the dense residual disk
dominates; its shape changes from quasi-elliptical to near
circular; the second arm of spiral shock forms.

The Figure 4 also presents the results of run \lq\lq A" and shows the
density distributions over the equatorial plane for other 6
moments of time $t_0+17.64P_{orb}$, $17.67P_{orb}$,
$17.70P_{orb}$, $17.73P_{orb}$, $17.76P_{orb}$, $17.79P_{orb}$,
covering the full period of the blob revolution for the stage
when the new quasi-steady-state solution is established.  The
maximal density in Fig.~3 corresponds to
$\rho\simeq0.02\rho(L_1)$. The analysis of these drawings shows
that obtained flow structure is similar to the solution for the
case of constant rate of mass transfer (see, e.g., Bisikalo et
al. 1998$^{\cite{dima98}}$, 2000$^{\cite{petr}}$, Paper I). It
is seen that the stream from $L_1$ dominates. It is also seen
that the uniform morphology of the system stream--disk results
in quasi-elliptical shape of the disk and the absence of `hot
spot' in zone of stream--disk interaction ($X\simeq0.8A$,
$Y\simeq-0.1A$).  At the same time the interaction of the gas of
circumbinary envelope with the stream results in the formation
of an extended shock wave located along the stream edge (`hot
line'). The Figure~4 also manifests the formation of only one
arm of tidally induced spiral shock located in I and II
quadrants of coordinate plane.  The flow structure in the place
where the second arm should be is defined by the stream from
$L_1$ which appears to suppress the formation of another arm of
the tidally induced spiral shock.

\begin{figure*}[p]
\centerline{
\hbox{\psfig{figure=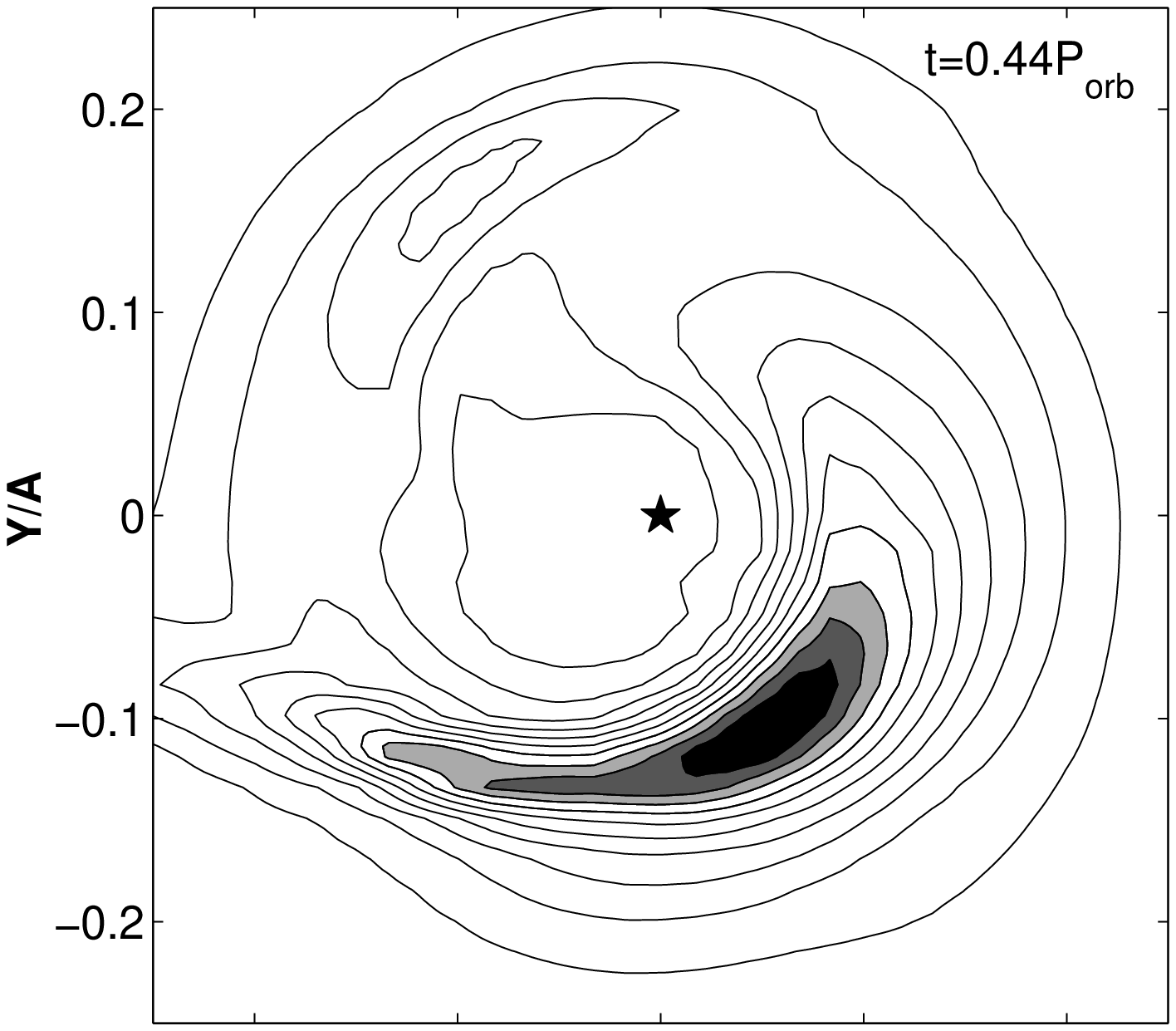,width=80mm}}
\hbox{\psfig{figure=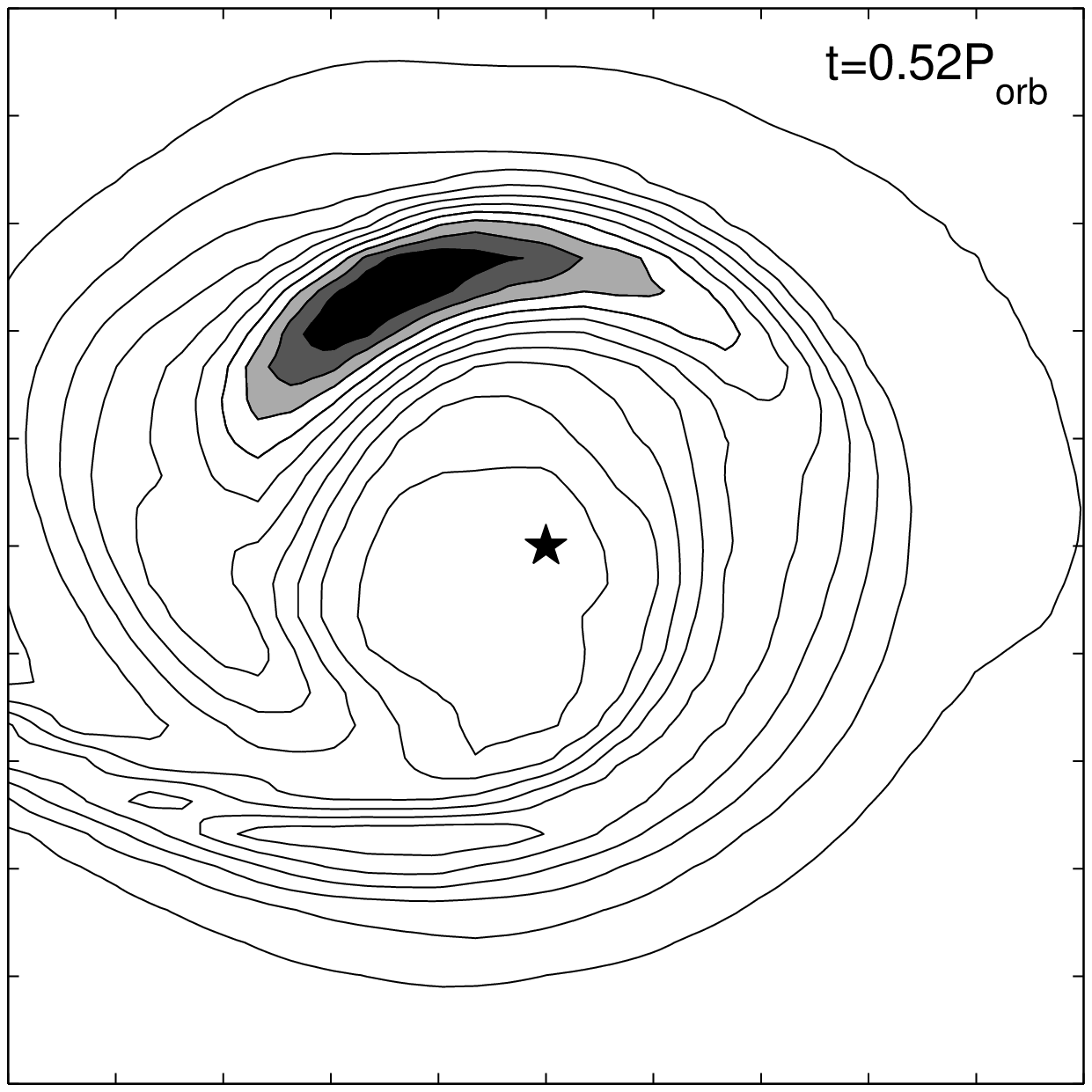,width=71mm}}
}
\centerline{
\hbox{\psfig{figure=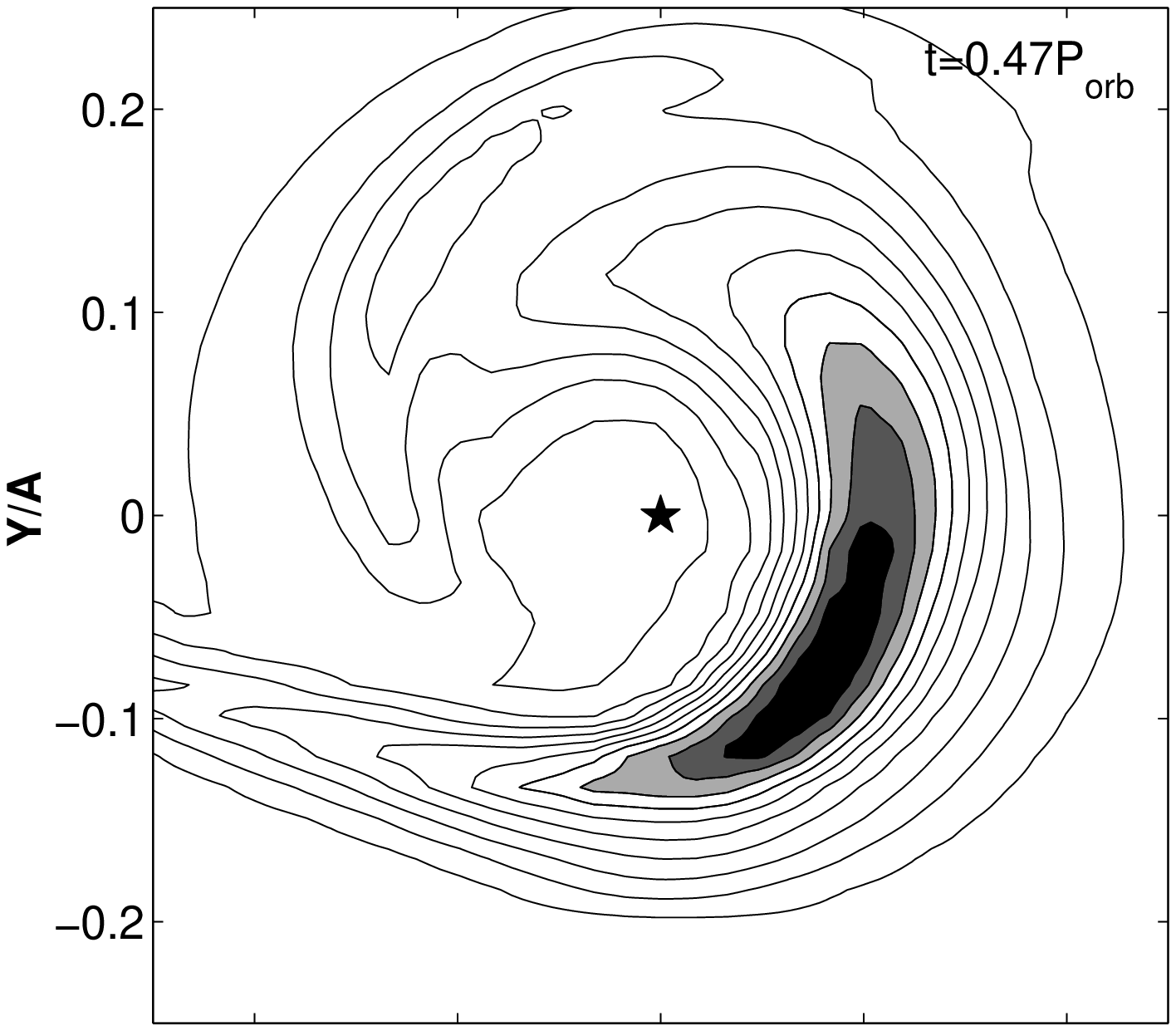,width=80mm}}
\hbox{\psfig{figure=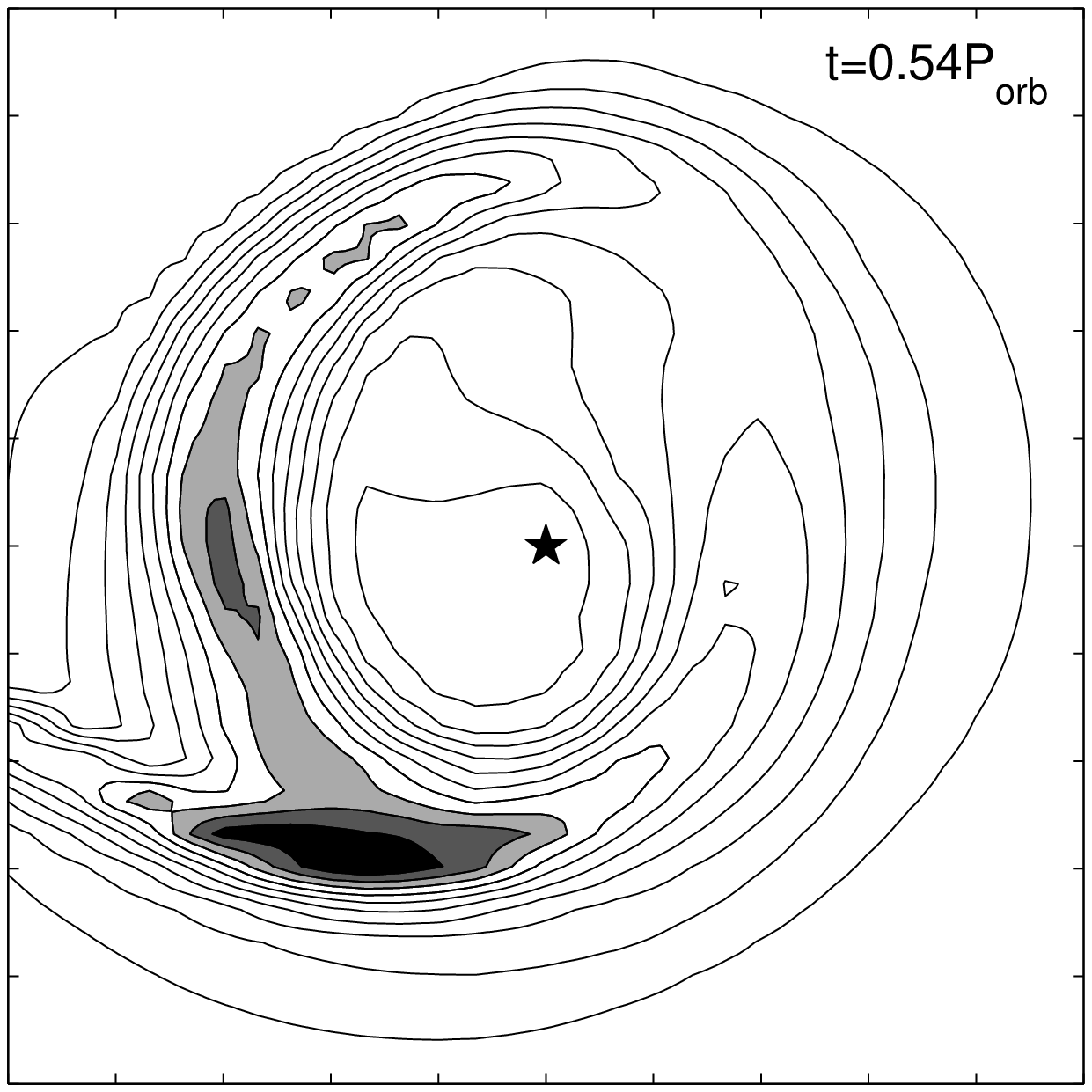,width=71mm}}
}
\centerline{
\hbox{\psfig{figure=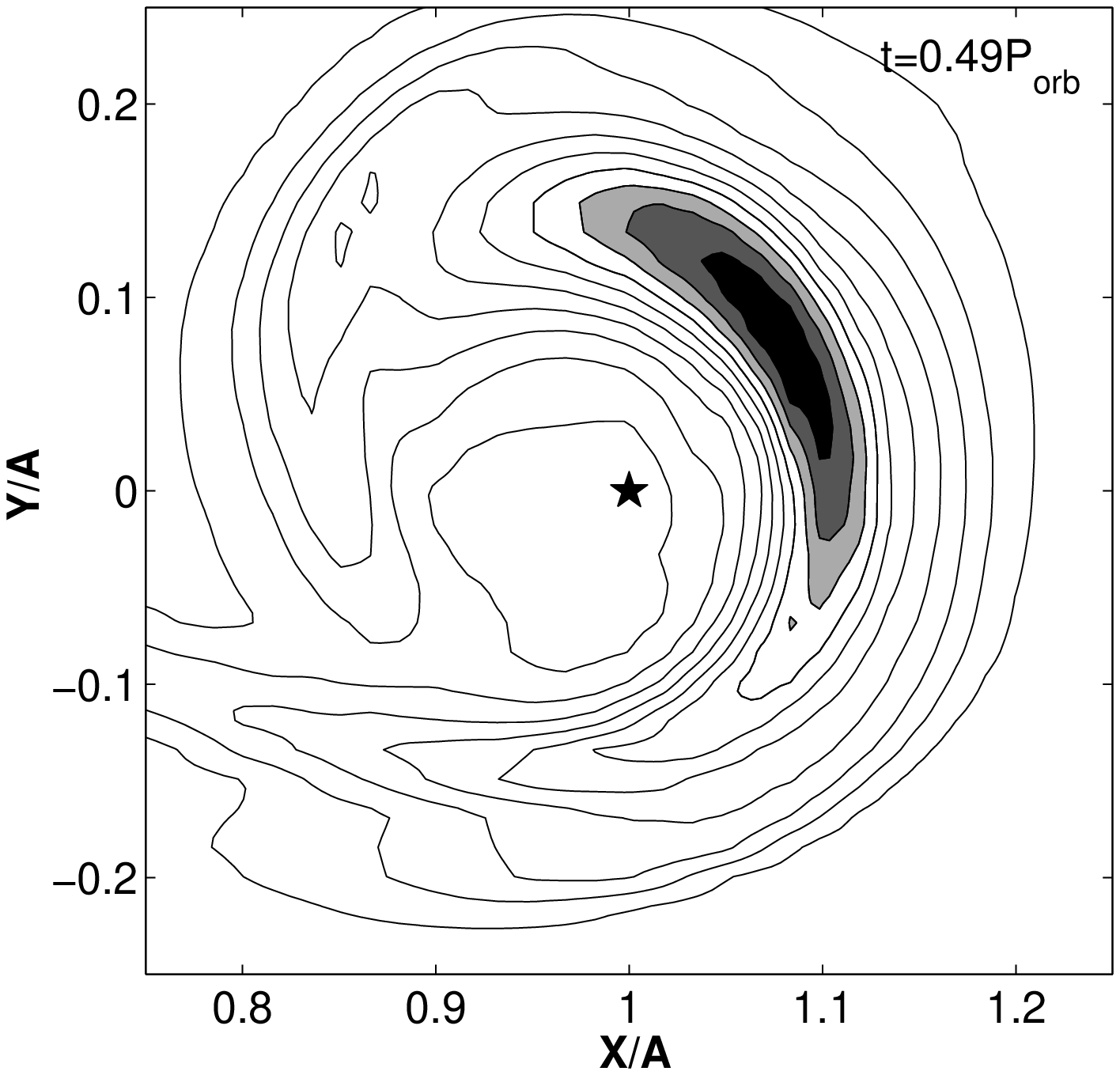,width=80mm}}
\hbox{\psfig{figure=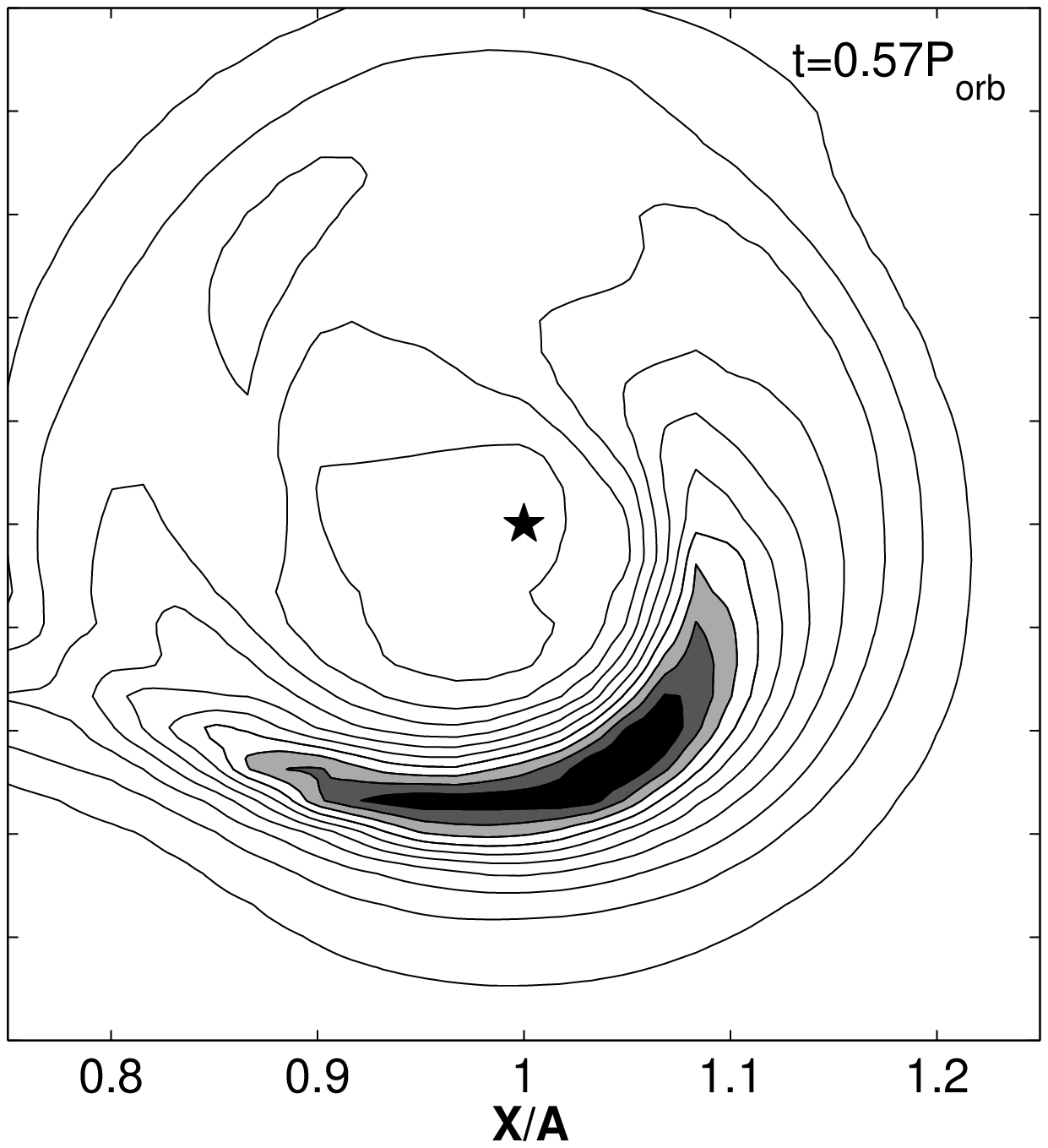,width=71mm}}
}
\caption{\footnotesize Density distribution over the equatorial
plane for run \lq\lq A". Results are presented for the moments of time
$t_0+0.44P_{orb}$, $0.47P_{orb}$, $0.49P_{orb}$, $0.52P_{orb}$,
$0.54P_{orb}$, $0.57P_{orb}$, covering the full period of the
blob revolution for the transient stage. The maximal density
corresponds to $\rho\simeq0.035\rho(L_1)$.}
\end{figure*}

\begin{figure*}[p]
\centerline{
\hbox{\psfig{figure=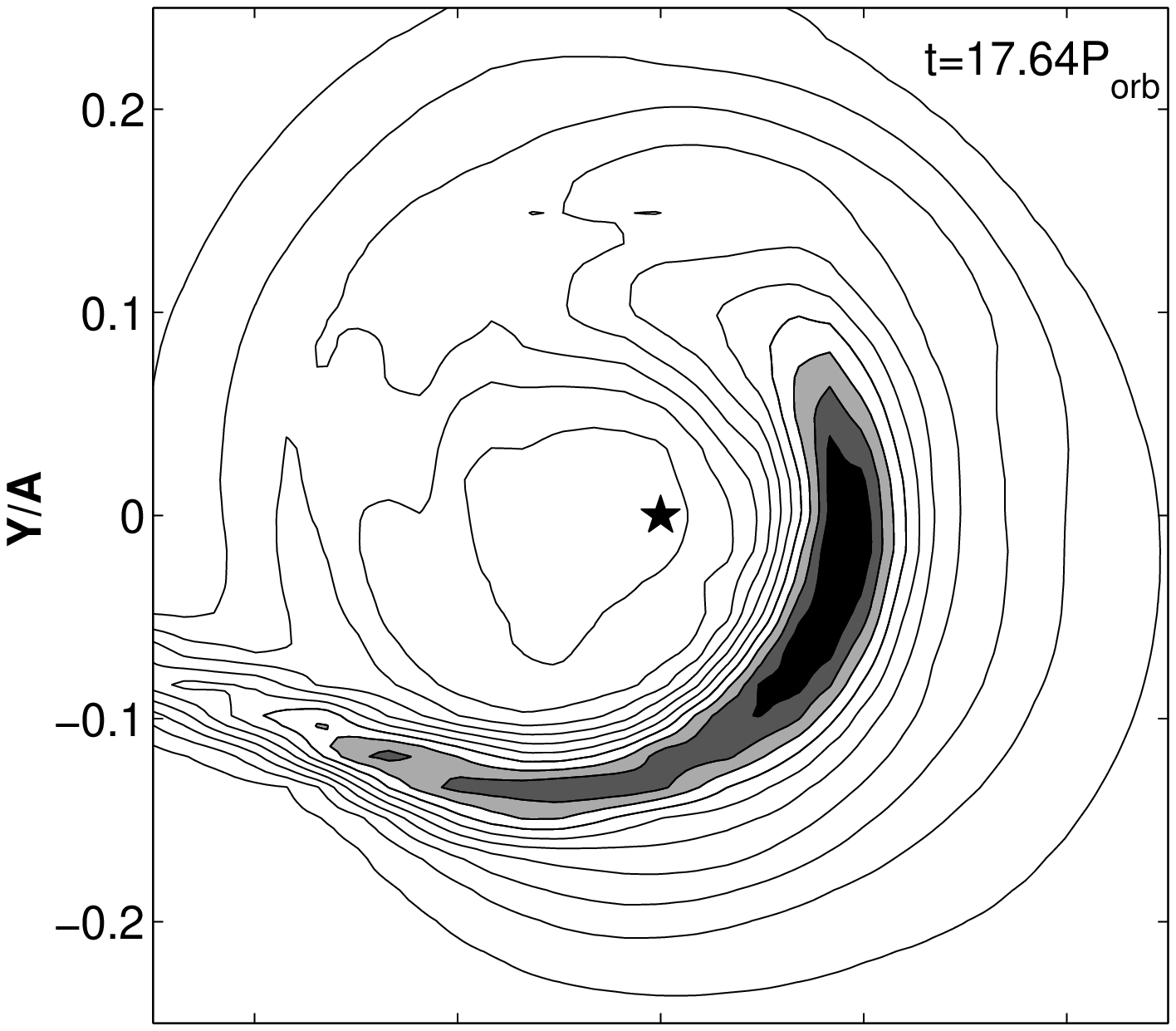,width=80mm}}
\hbox{\psfig{figure=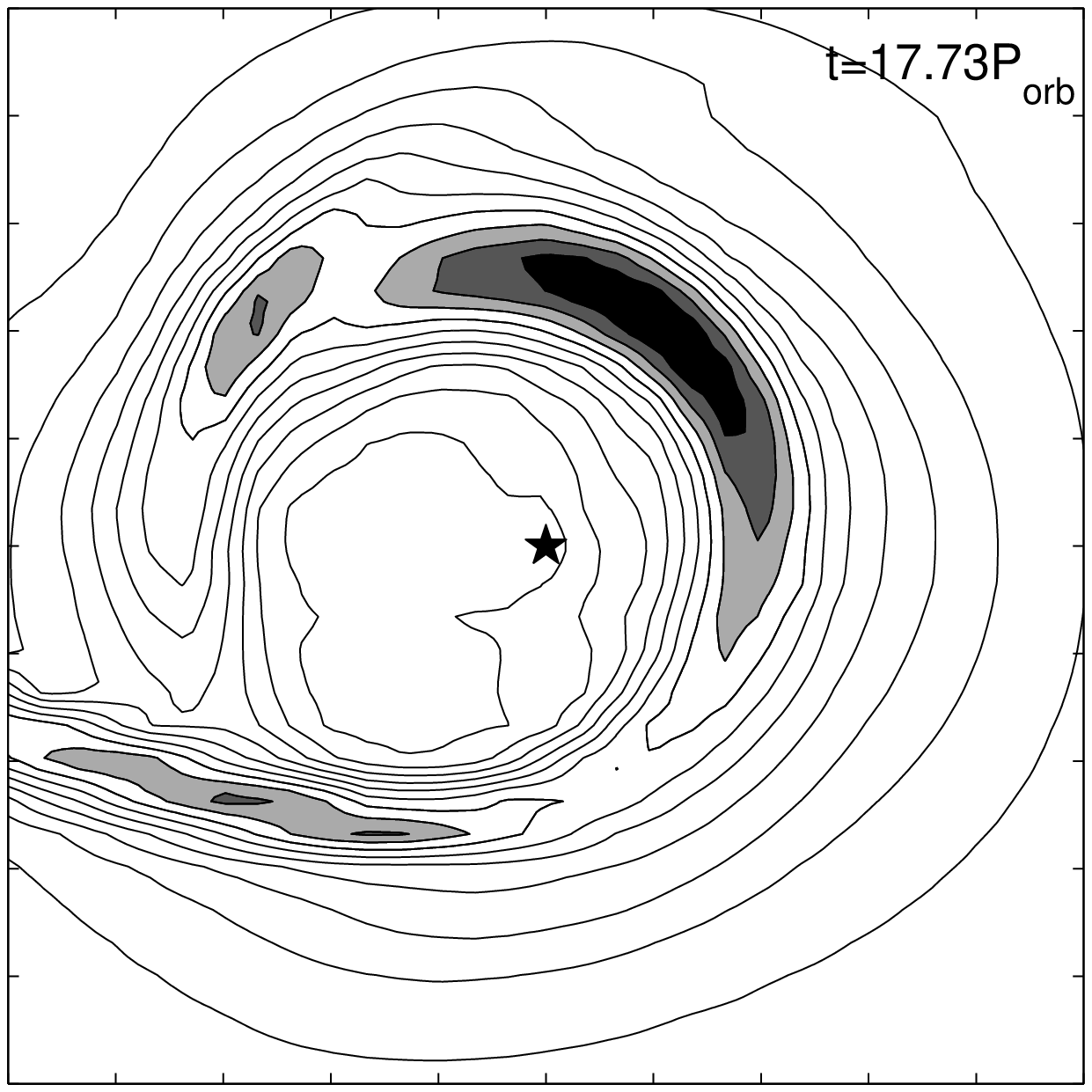,width=71mm}}
}
\centerline{
\hbox{\psfig{figure=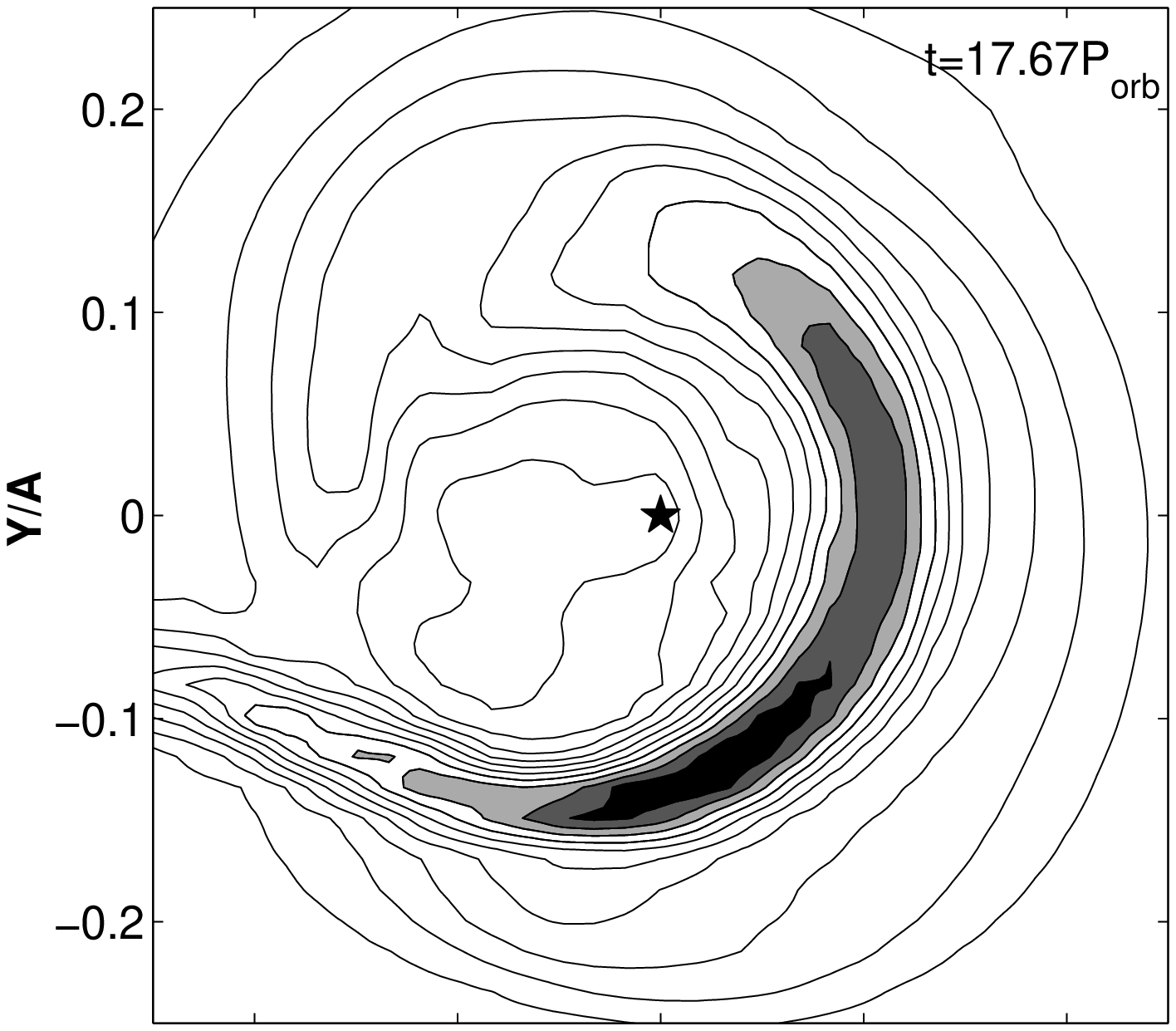,width=80mm}}
\hbox{\psfig{figure=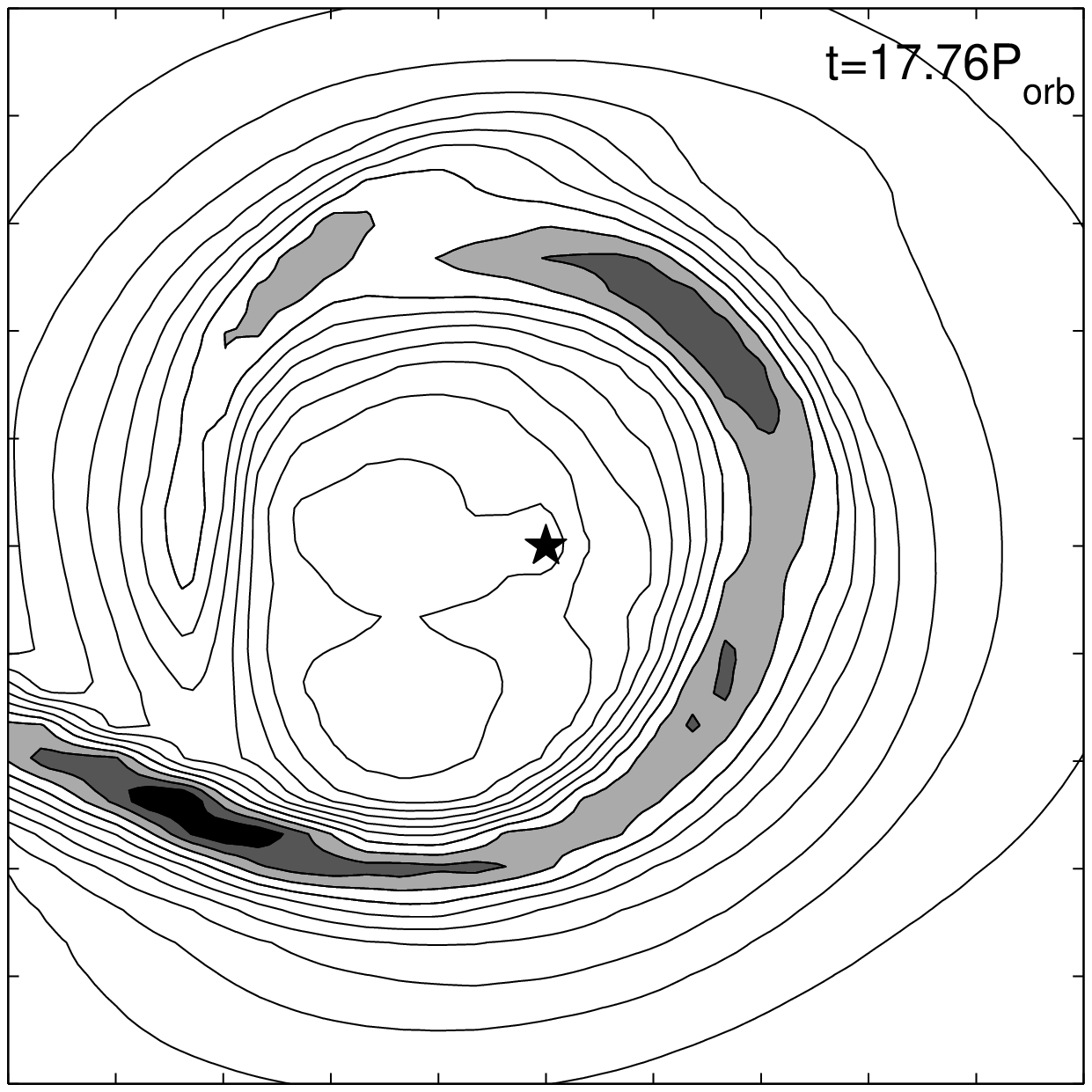,width=71mm}}
}
\centerline{
\hbox{\psfig{figure=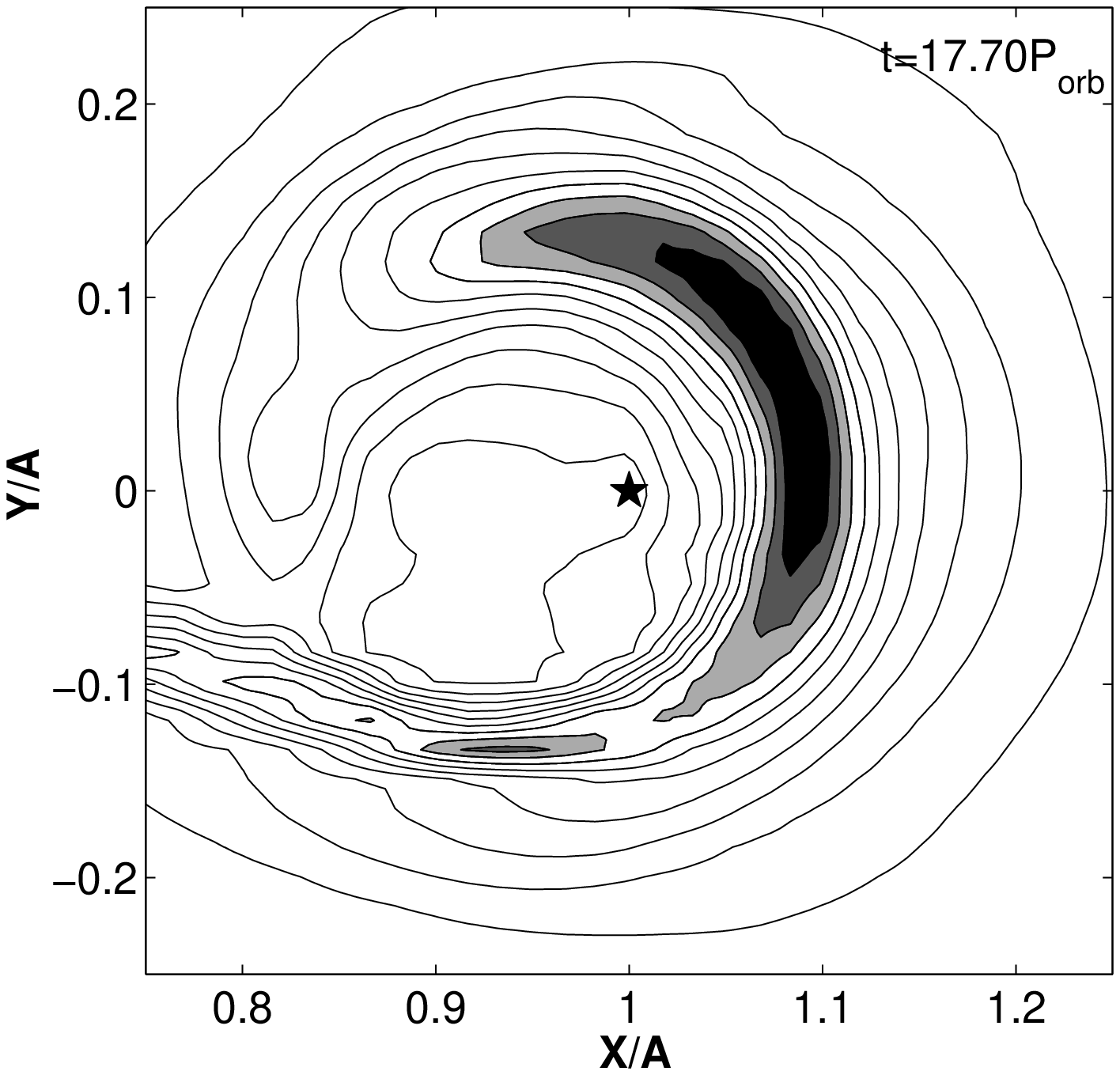,width=80mm}}
\hbox{\psfig{figure=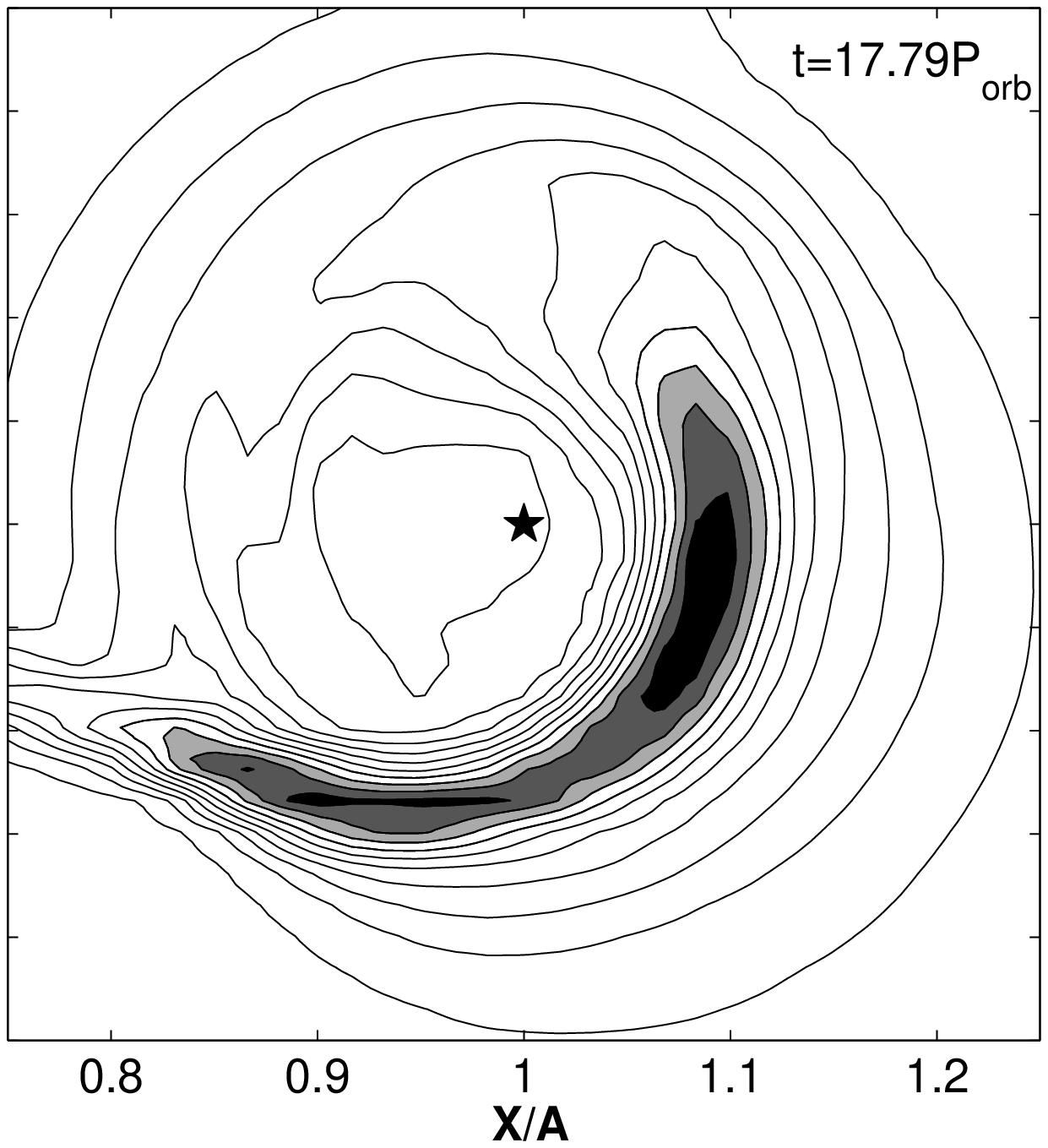,width=71mm}}
}
\caption{\footnotesize The same as in Fig.~3 for other six
moments of time:  $t_0+17.64P_{orb}$, $17.67P_{orb}$,
$17.70P_{orb}$, $17.73P_{orb}$, $17.76P_{orb}$, $17.79P_{orb}$,
covering the full period of the blob revolution for the stage
when the new quasi-steady-state solution is established.  The
maximal density corresponds to $\rho\simeq0.02\rho(L_1)$.}
\end{figure*}

Nevertheless, opposite to the case of constant rate of mass
transfer the blob remains to live and gives the periodic changes
of disk structure. The Figure~4 shows that the blob tends to
spread uniformly but passing through the arm of spiral shock it
is retarded, this leads to the formation/sustaining of compact
blob. This also leads to periodic changes of disk structure in
a new quasi-steady-state solution. The Fourier analysis shows
(see Fig.~2a) that for run \lq\lq A" dominating harmonics are
$0.155P_{orb}$ and $0.165P_{orb}$.  The beating of these two
gives long-period oscillations with period $\sim 5.1P_{orb}$.
First (short-period) harmonic appears to be due to interaction
of the blob with the arm of spiral shock and the second harmonic
is due to interaction of peripherical parts of the blob with the
stream of matter from $L_1$, the period of latter harmonic being
larger. The larger period is result of differential rotation of
the disk -- peripherical parts of the blob locate at larger
radii and rotate slower. After interaction with the stream this
lag decreases and the blob becomes compact.

Comparative analysis of Figs~1--4 shows that the mean period of
the blob rotation is the same both for transient stage when two
arms of spiral shock exist and for new quasi-steady-state
solution with one arm of spiral shock. At the same time the
density contrast between the disk and the blob changes
significantly due to long-period modulations (Fig.~5). It is
seen that on the transient stage when the dense disk dominates
the contrast is $\sim1.5$. Later, after reaching the recurrent
quasi-steady-state solution with two dominating harmonics and
long-period modulations the density contrast varies from
$\sim1.1$ to $\sim1.8$.

\begin{figure}[t]
\centerline{\hbox{\psfig{figure=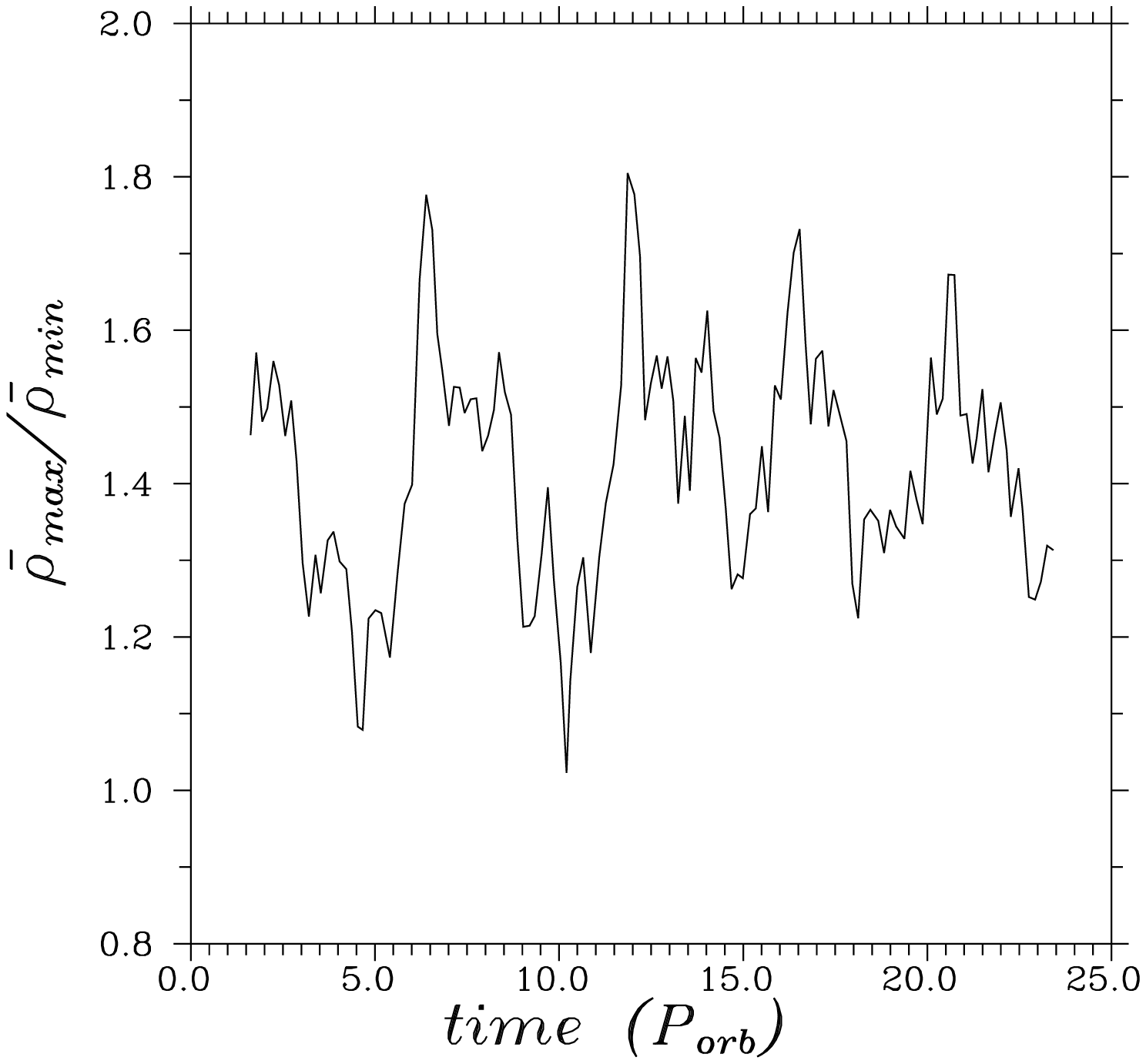,width=7cm}}}
\caption{\footnotesize The density contrast between the disk and
the blob versus time.}
\end{figure}

\begin{figure*}[p]
\centerline{
\hbox{\psfig{figure=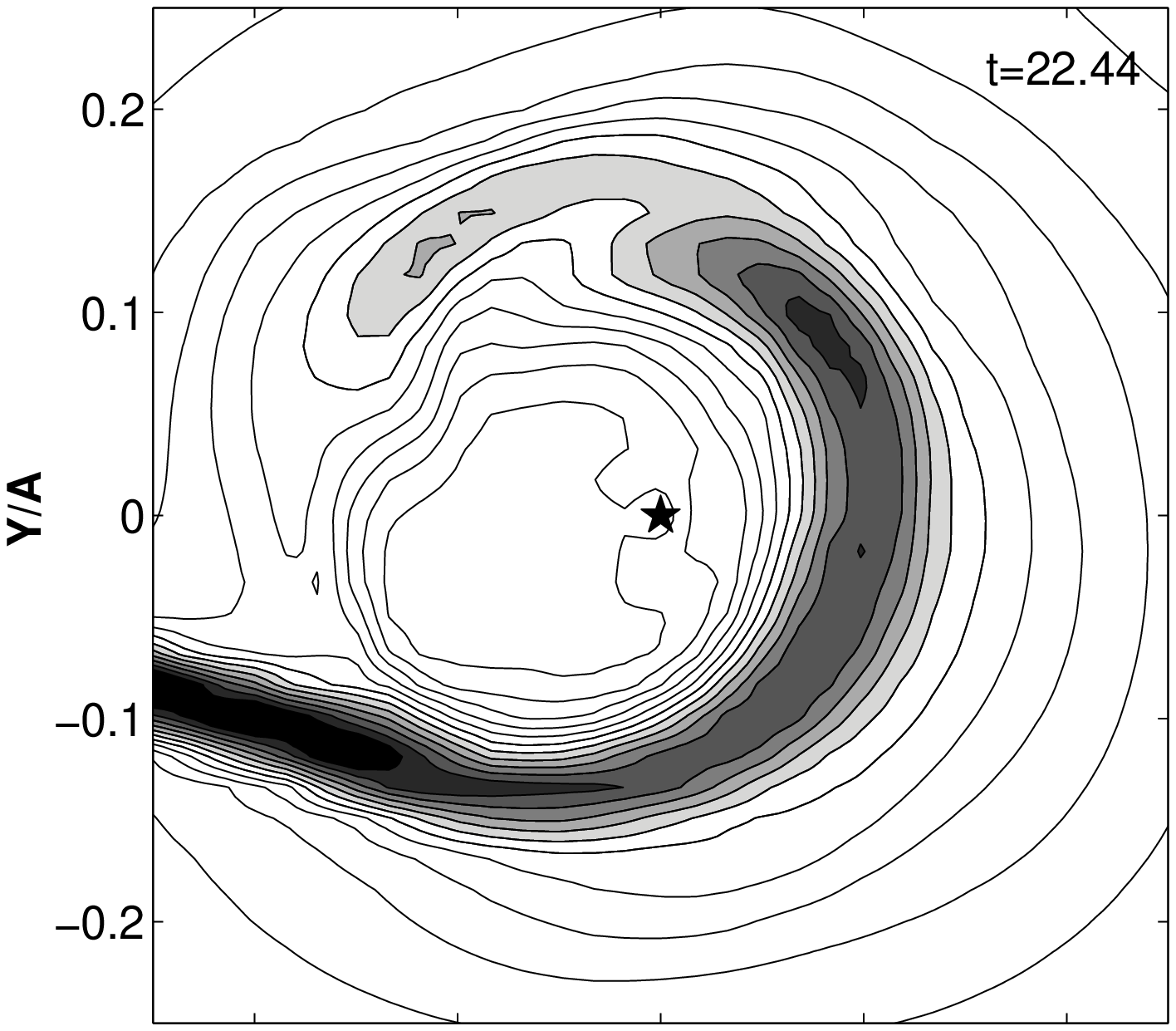,width=80mm}}
\hbox{\psfig{figure=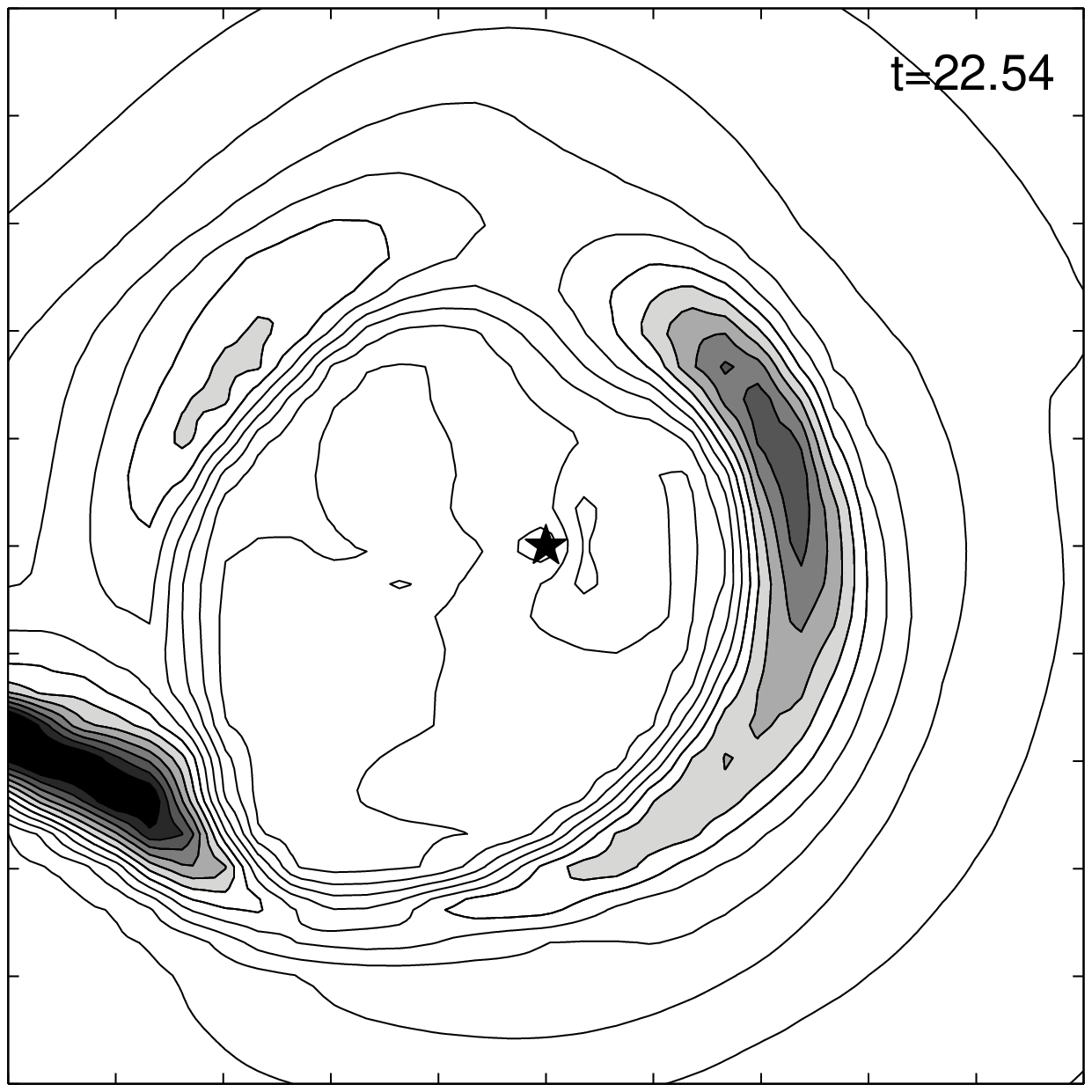,width=71mm}}
}
\centerline{
\hbox{\psfig{figure=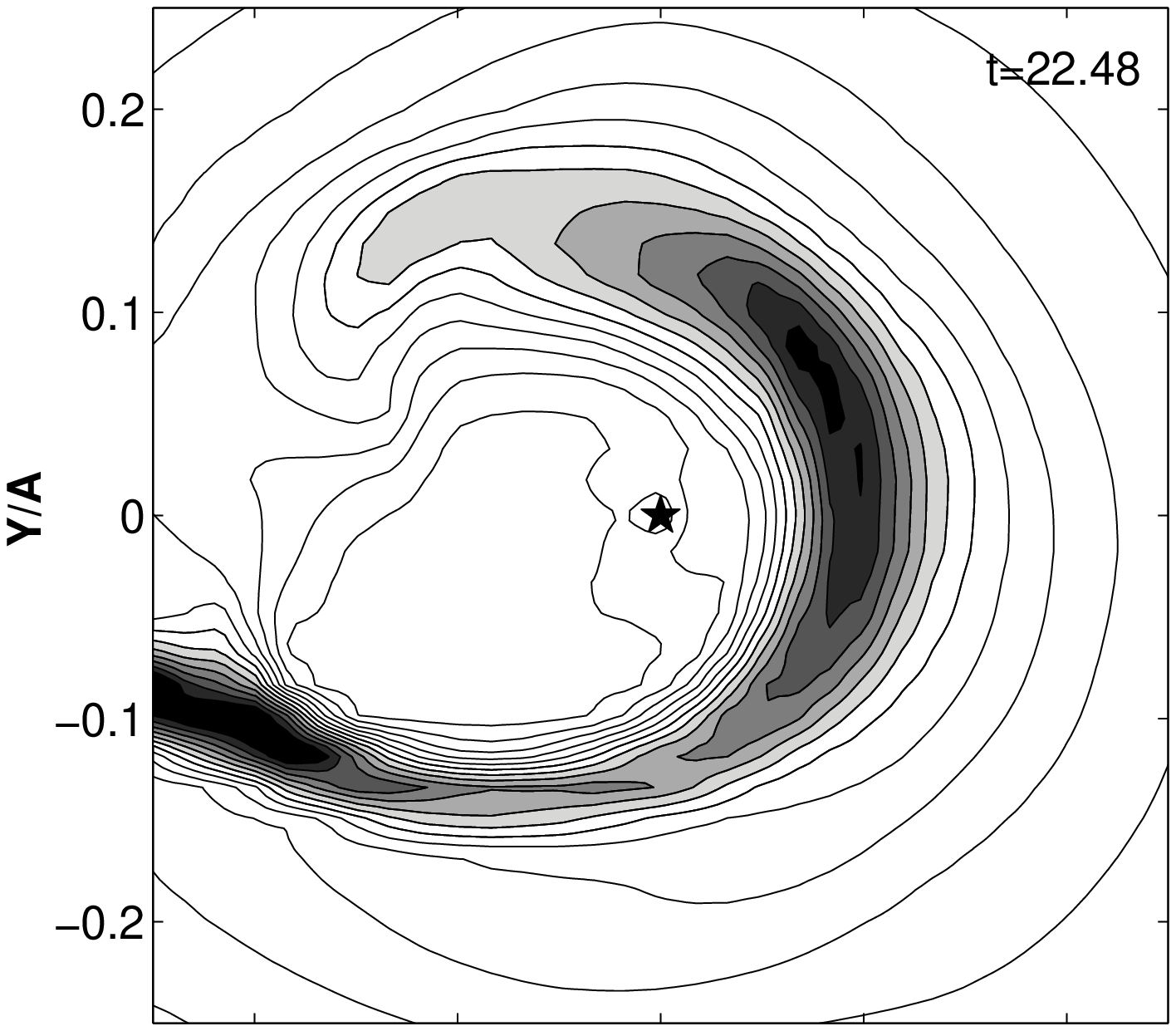,width=80mm}}
\hbox{\psfig{figure=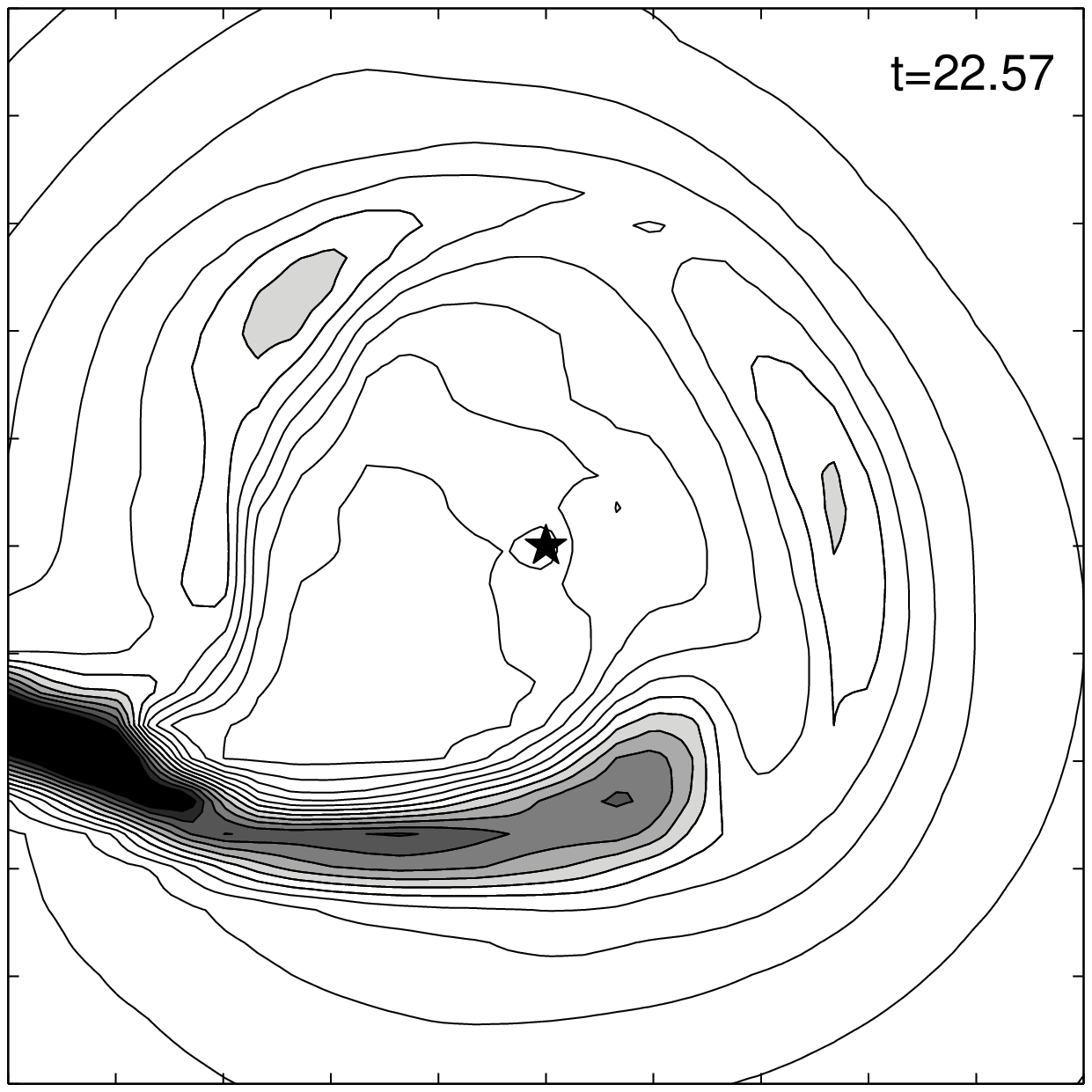,width=71mm}}
}
\centerline{
\hbox{\psfig{figure=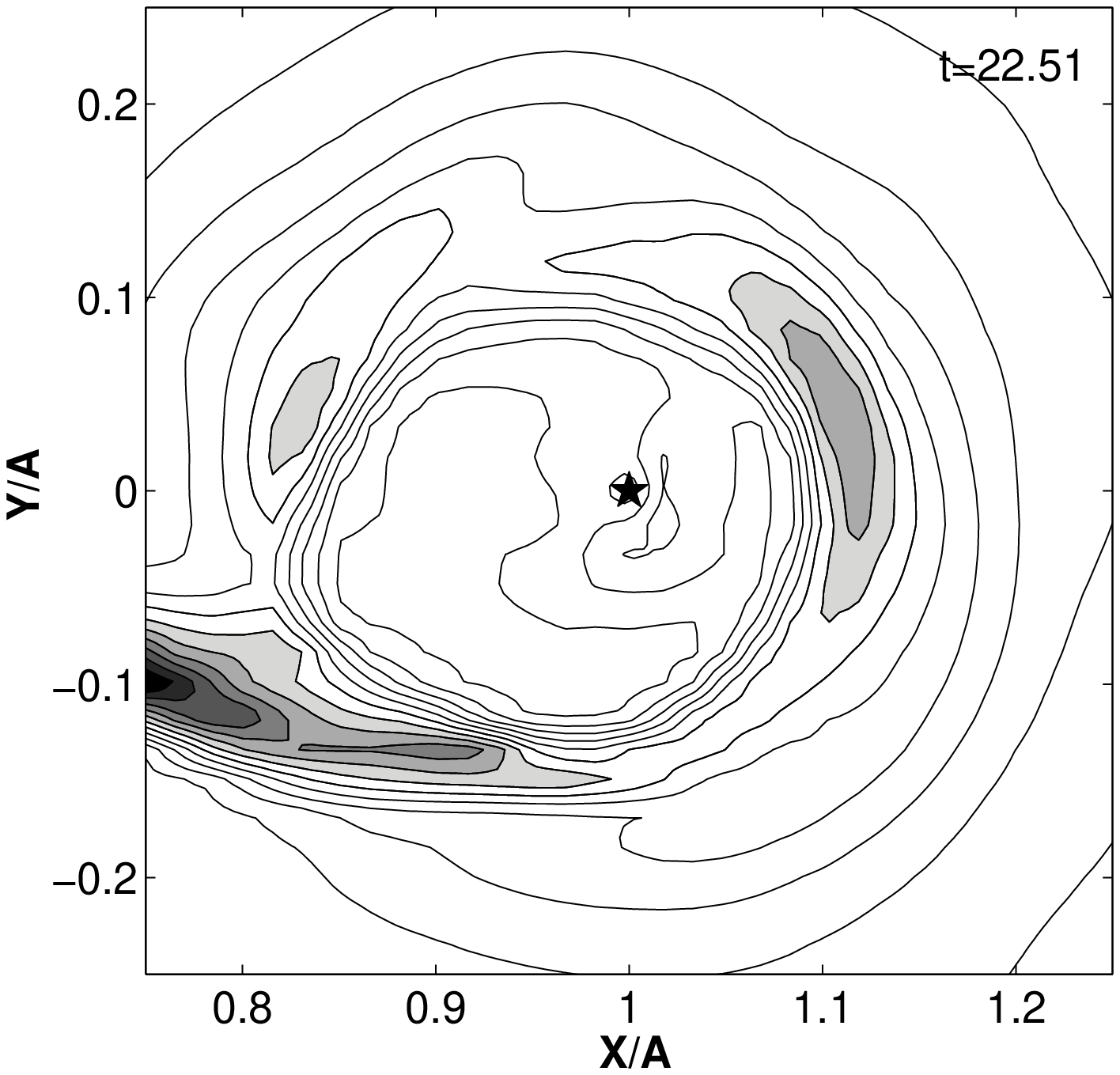,width=80mm}}
\hbox{\psfig{figure=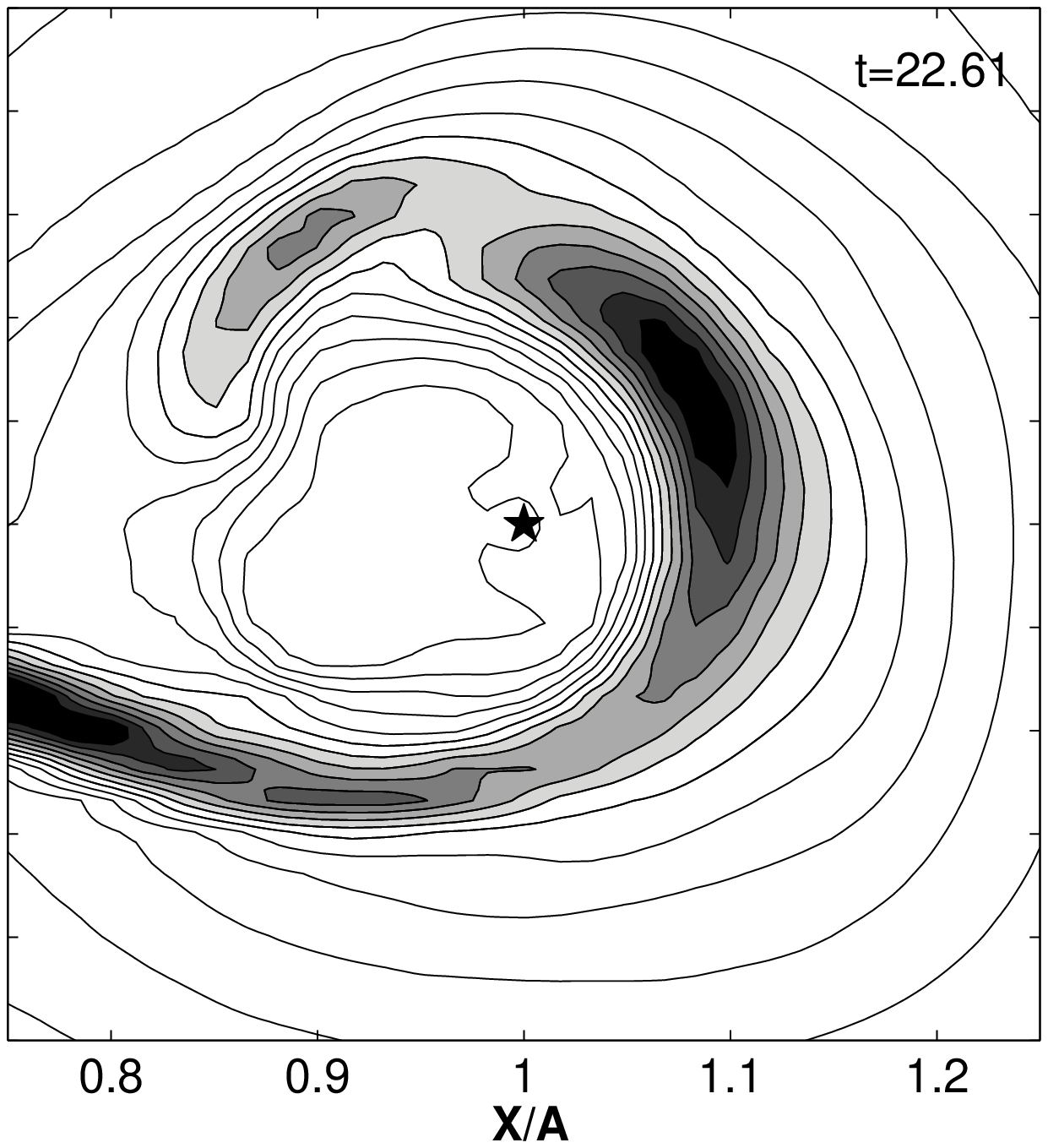,width=71mm}}
}
\caption{\footnotesize Density distribution over the equatorial
plane for run \lq\lq B". Results are presented for the moments of time
$t_0+22.45P_{orb}$, $22.48P_{orb}$, $22.51P_{orb}$,
$22.54P_{orb}$, $22.57P_{orb}$, $22.61P_{orb}$, covering the
full period of the blob revolution for the new
quasi-steady-state stage of run \lq\lq B".  The maximal density
corresponds to $\rho\simeq0.0035\rho(L_1)$.}
\end{figure*}

Resuming the results of run \lq\lq A", we can conclude that the
solution at the recurrent quasi-steady-state stage is defined by
the stream from $L_1$ and the presence of the blob manifests in
periodic variations of the disk structure.

Let us consider now the results of run \lq\lq B" when the mass
transfer rate was decreased 10 times and the gas of the stream
has significantly less density. Similar to run \lq\lq A" on the
transient stage the dense disk and the blob dominate. But
opposite to run \lq\lq A" the blob continues to dominate even at the
recurrent quasi-steady-state stage. Opposite to run \lq\lq A" the blob
changes the general flow structure, in particular, the blob
`blows off' the stream when reaching it.  This is seen from
Fig.~6 where the density distributions over the equatorial plane
for moments of time $t_0+22.45P_{orb}$, $22.48P_{orb}$,
$22.51P_{orb}$, $22.54P_{orb}$, $22.57P_{orb}$, $22.61P_{orb}$,
covering the full period of the blob revolution for the new
quasi-steady-state stage of run \lq\lq B".  The maximal density in
Fig.~6 corresponds to $\rho\simeq0.0035\rho(L_1)$. The small
value of density confirms the significant influence of the blob
on the flow structure on all stages of run \lq\lq B". The Fourier
analysis of results of the run \lq\lq B" shows (see Fig.~2b) that
dominating harmonics are $0.18P_{orb}$ and $0.21P_{orb}$. The
beating of these two gives long-period oscillations with period
$\sim 2.5P_{orb}$.  Similar to run \lq\lq A" these harmonics appears
to be due to interaction of the blob with the arm of spiral
shock and with the stream of matter from $L_1$.

The analysis of these results as well as the results of Paper I
shows that for binaries with spiral shocks small disturbances
resulting from variations of mass transfer rate lead to
formation of the blob. Moreover, on the dynamical timescale the
blob is supported by the presence of spiral shocks and this
mechanism is independent on the variations of mass transfer
rate. Thus, we can conjecture that not only mass transfer rate
variation but every disturbance of disk structure will transform
into the blob. The lifetime of the blob is determined by the
lifetime of spiral shock which prevents the blob spreading in
dynamical timescale, and dissipation which governs the blob
spreading in viscous timescale.

\section{The blob parameters and the variability of light curves
of CVs with spiral shocks}

Spiral shocks or, more exactly, the dense post-shock regions
attached to the arms of the spiral shock were often considered
as probable source of variability of astrophysical objects (see,
e.g., Chakrabarti \& Wiita 1993a$^{\cite{chak93a}}$,
1993b$^{\cite{chak93b}}$). Recently we argued in Paper I that the
dense blob can detach from the front of spiral shock and further
travel through the disk. In Paper I as well as in the present
work the blob is formed as a result of disturbance of the disk
structure due to variation of mass transfer rate. Once forming
blob tends to distribute uniformly along the disk due to
differential rotation but retards after passing trough spiral
shock so the compactness of the blob retains/resumes. Later the
increasing density contrast between the blob and the disk as
well as growths of pressure gradient will force the standstilled
blob to move. When reaching again the front of spiral shock the
process of formation/retainment of the blob is repeated. Results
of 3D gasdynamical calculations show that the blob travels
through the disk during tens of orbital periods, its main
characteristics being conserved. The blob ought spread under the
action of dissipation but the viscous timescale is much greater
than the time of our calculations so we can neglect the viscous
spreading of the blob. We should like to stress that the we have
chosen the duration of our calculations (up to $\sim25P_{orb}$)
not arbitrary but in agreement with typical length of the
outburst decline (tens of $P_{orb}$). It is on the stage of
outburst decline the decreasing of mass transfer rate is
possible, and, besides, on this stage the spiral shocks are
observable (or, more exactly, the most reasonable assumptions on
their existence can be made here).

To determine the typical features of the blob in CVs it is
necessary to conduct a set of calculations covering all the
range of parameters which are relevant to CVs. Nevertheless,
based on the results of Paper I as well as on the results of
this work we can conclude that generally the period of the blob
revolution depends on the value of viscosity: it equals to
$0.15\div0.20P_{orb}$ for $\alpha\sim0.05$, and to
$0.10\div0.15P_{orb}$ for $\alpha\sim0.01$. This means that for
typical accretion disk with $\alpha\sim0.01\div0.1$ the period
of the blob revolution should be in range $0.10\div0.25P_{orb}$.
Another important property of the blob is the density contrast
w.r.t. the rest of the disk since this parameter determines the
observational evidences of the blob. As it was shown above the
value of density contrast is independent on the rate of mass
transfer and changes periodically around the mean value of
$\sim1.5$. The nature of mechanism of formation/sustaining of
the blob permits us to believe that this value is typical for
different CVs.

Let us consider the virtual observation evidences of the blob
and compare it with the real observations. The existence of the
blob is maintained by spiral shocks, so we have chosen the
observation of IP Peg which was gave the first evidences of the
spiral shocks (Steeghs, Harlaftis \& Horne
1997$^{\cite{steeghs97}}$; Harlaftis et al.
1999$^{\cite{harlaftis99}}$) and EX Dra where the existence of
spiral shocks is also probable (Baptista \& Catal\'an
1999$^{\cite{baptista990}}$, 2000$^{\cite{baptista99}}$;
Joergens et al.  2000$^{\cite{joergens2000a}}$; Joergens, Spruit
\& Rutten 2000$^{\cite{joergens2000b}}$).

Since the blob is preserved from the spreading by the existence
of spiral shock let us consider the link between the light curve
variations and the appearance/disappearance of spiral shocks.
The standard treating of IP Peg and EX Dra suggests the
extension of the accretion disk and formation of spiral shocks
in it at some stage. After the outburst ends and the quiescence
state establishes the radius decreases so tidal interaction
can't induce spiral shocks anymore. Thus, the light curves
should manifest quasi-periodic variations only during outburst
and, possible, some time after it.\footnote{Note that if the
viscosity in the disk is small then the dissipative spreading is
negligible and the blob can exist even when the spiral shocks
vanish. Correspondingly, the light curve variations can be
observable after that while with decreasing magnitude.}
Observations show variations of uneclipsed parts of the light
curves for both systems during outburst and immediately after
it. The width of these variations is $\sim0^{\rm
P}\!\!.1\div0^{\rm P}\!\!.2$ for EX Dra (Baptista, Catal\'an \&
Costa 2000$^{\cite{baptista2000}}$) and for IP Peg (Marsh \&
Horne 1990$^{\cite{marsh90}}$; Webb et al.
1999$^{\cite{webb99}}$), the amplitude of variations
sufficiently decreasing when the quiescence state establishes.
These observations confirm the proposed model on the qualitative
level: it is seen that variations of uneclipsed parts of light
curves correlate with the presence of spiral shocks, and the
width of variations corresponds to the theoretical one.

As another test for the suggested model let us consider the
quantitative evidences of the blob existence. The analysis of
light curves shows that the disk deposits the major fraction of
luminosity during outburst (up to $\slantfrac{5}{6}$ for IP Peg,
see, e.g., Khruzina et al. 2001$^{\cite{tanya2001}}$), so the
amplitude of light curve variations should be approximately
equal to the density contrast. Our calculations give the mean
value of density contrast as 1.5 so taking into account the
relative luminosity of the disk we can expect the amplitude of
variations on light curves as $\sim 40$\%. Observations show the
amplitude of variations of the flux on uneclipsed parts of light
curves as $\sim 30$\% for IP Peg (see, e.g., Steeghs et al.
1996$^{\cite{steeghs96}}$; Morales-Rueda, Marsh \& Billington
2000$^{\cite{luisa}}$) and $\sim 15$\% for EX Dra (Baptista,
Catal\'an \& Costa 2000$^{\cite{baptista2000}}$). Unfortunately,
the analysis of light curve variations in quiescence is
inconvenient due to both decreasing of density contrast and
relative luminosity of the disk ($\slantfrac{1}{5}$ for IP Peg,
see, e.g., Khruzina et al. 2001$^{\cite{tanya2001}}$), so we
restrict ourself by consideration of quantitative estimations
for outburst only. The comparison of these estimations confirms
the proposed model on the quantitative level as well.

As an additional test of the model applicability we would like
to stress its relevance to the explanation of variation of
emission lines profiles in CVs. For example, the observations of
$H_\alpha$ profiles in the disk of binary SS Cyg
(Mart{\'\i}nez-Pais et al. 1994$^{\cite{franco}}$) show
its essentially asymmetrical nature, the distance between peaks
being significantly variable. The presence of spiral shocks
permits to explain the asymmetry of line profiles (Chakrabarti
\& Wiita 1993a$^{\cite{chak93a}}$), but the spiral shocks by
itself can't give the mechanism resulting in variable distance
between peaks. The mechanism suggested in this work do explain
both the asymmetry of line profiles (in the same manner as
above) and the variability of the distance between peaks (by the
variability of the projection of blob velocity). Note also that
the sporadic low-magnitude variations on the light curve
(flickering) can also be a consequence of the presence of the
blob in the disk. Disturbances of the disk structure due to
periodic travelling of the blob as well as due to stream/blob
interaction can result in low-amplitude oscillations manifesting
as sporadic variations on the light curve.

\section{Conclusions}

Light curves of CVs show periodic or quasi-periodic photometric
modulations with typical width $\sim0^{\rm P}\!\!.1\div0^{\rm
P}\!\!.2$. The 3D gasdynamical simulations of flow structure in
binaries disturbed by the mass transfer rate variation prove the
appearing a new dense formation -- blob, the latter moving
through the disk with variable velocity. Long lifetime of the
blob, significant density contrast between the blob and the rest
of accretion disk (the mean value is $\sim 1.5$), and
characteristic period of its revolution $\sim0.1\div0.2P_{orb}$
permit us to consider this formation as a possible reason for
observable quasi-periodic variations of light curves of CVs.
Note that if suggested mechanism is valid then the variations of
light curves of CVs give an additional proof for the spiral
shocks existence.

\section*{Acknowledgments}

The work was partially supported by Russian Foundation for Basic
Research (projects NN 99-02-17619, 00-01-00392) and by grants of
President of Russia (99-15-96022, 00-15-96722). Authors wish to
thank A.M.Fridman and O.V.Khoruzhii for useful discussions.

\clearpage

\onecolumn


\begin{thebibliography}{9999}

%\raggedright

\footnotesize

\bibitem{baptista990}
Baptista~R., Catal\'an~M.S. 1999, Changes in the Structure of
the Accretion Disc of HS1804+67 through the Outburst Cycle, in
\lq\lq Cataclysmic Variables: a 60th Birthday Symposium in Honour of
Brian Warner", eds P.Charles et al., New Astron. Rev., in press
(preprint astro-ph/9905096)

\bibitem{baptista99}
Baptista~R., Catal\'an~M.S. 2000, Eclipse Studies of the Dwarf
Nova EX Dra, {\it Astrophys. J.}, {\bf 316}, L529

\bibitem{baptista2000}
Baptista~R., Catal\'an~M.S., Costa~L. 2000, Eclipse Studies of
the Dwarf Nova EX Draconis, {\it Monthly Notices Roy. Astron.
Soc.}, {\bf 316}, 529

\bibitem{dima98}
Bisikalo~D.V., Boyarchuk~A.A., Chechetkin~V.M., Kuznetsov~O.A.,
Molteni~D. 1998, 3D Numerical Simulation of Gaseous Flows
Structure in Semidetached Binaries, {\it Monthly Notices Roy.
Astron. Soc.}, {\bf 300}, 39

\bibitem{petr}
Bisikalo~D.V., Harmanec~P., Boyarchuk~A.A., Kuznetsov~O.A.,
Hadrava~P. 2000, Circumstellar Structures in the Eclipsing
Binary $\beta$ Lyr A. Gasdynamical modelling confronted with
observations, {\it Astron. Astrophys.}, {\bf 353}, 1009

\bibitem{blob1}
Bisikalo~D.V., Boyarchuk~A.A., Kilpio~A.A., Kuznetsov~O.A.,
Chechetkin~V.M. 2001, Structure of Gaseous Flows in Semidetached
Binaries after Mass Transfer Termination, {\it Astron. Zh.}, in
press (preprint astro-ph/0102241), \underline{Paper~I}

\bibitem{chak93a}
Chakrabarti~S.K., Wiita~P.J. 1993a, Effects of Spiral Shocks on
Disk Emission Lines, {\it Astron. Astrophys.}, {\bf 271}, 216

\bibitem{chak93b}
Chakrabarti~S.K., Wiita~P.J. 1993b, Spiral Shocks in Accretion
Disks as a Contributor to Variability in Active Galactic Nuclei,
{\it Astrophys. J.}, {\bf 411}, 602

\bibitem{osher85}
Chakravarthy~S.R., Osher~S. 1985, A New Class of High Accuracy
TVD Schemes for Hyperbolic Conservation Laws, {\it AIAA Pap.},
N~85-0363

\bibitem{ladous93}
Hack~M., La~Dous~C. 1993, Cataclysmic Variables and Related
Objects. Washington: US Gov. Printing Office

\bibitem{harlaftis99}
Harlaftis~E.T., Steeghs~D., Horne~K., Mart{\'\i}n~E.,
Magazz\'u~A. 1999, Spiral Shocks in the Accretion Disc of IP Peg
during Outburst Maximum, {\it Monthly Notices Roy. Astron.
Soc.}, {\bf 306}, 348

\bibitem{joergens2000a}
Joergens~V., Mantel~K.H., Barwig~H., B\"arbanter~O., Fiedler~H.
2000, Reconstruction of Emission Sites in Dwarf Nova EX
Draconis, {\it Astron.  Astrophys.}, {\bf 354}, 579

\bibitem{joergens2000b}
Joergens~V., Spruit~H.C., Rutten~R.G.M. 2000, Spirals in the
Disk of EX Dra, {\it Astron. Astrophys.}, {\bf 356}, L33

\bibitem{tanya2001}
Khruzina~T.S., Cherepashchuk~A.M., Bisikalo~D.V.,
Boyarchuk~A.A., Kuznetsov~O.A. 2001, The Interpretation of Light
Curves of IP Peg in the Model of Shock-free Interaction between
the Gas Stream and the Disk, {\it Astron. Zh.}, in press

\bibitem{marsh90}
Marsh~T.R., Horne~K. 1990, Emission-line Mapping of the Dwarf
Nova IP Pegasi in Outburst and Quiescence, {\it Astrophys. J.},
{\bf 349}, 593

\bibitem{franco}
Mart{\'\i}nez-Pais~I.G., Giovanelli~F., Rossi~C., Gaudenzi~S.
1994, An Optical Time-resolved Spectroscopic Study of SS Cygni,
{\it Astron. Astrophys.}, {\bf 291}, 455

\bibitem{luisa}
Morales-Rueda~L., Marsh~T.R., Billington~I. 2000, Spiral
Structure in IP Pegasi: How Persistent it? {\it Monthly Notices
Roy. Astron. Soc.}, {\bf 313}, 454

\bibitem{roe86}
Roe~P.L. 1986, Characteristic-Based Schemes For the Euler
Equations, {\it Ann. Rev. Fluid Mech.}, {\bf 18}, 337

\bibitem{steeghs96}
Steeghs~D., Horne~K., Marsh~T.R., Donati~J.F. 1996, Slingshot
Prominences during Dwarf Nova Outburst? {\it Monthly Notices
Roy. Astron. Soc.}, {\bf 281}, 626

\bibitem{steeghs97}
Steeghs~D., Harlaftis~E.T., Horne~K. 1997, Spiral Structure in
the Accretion Disc of the Binary IP Pegasi, {\it Monthly Notices
Roy. Astron. Soc.}, {\bf 290}, L28

\bibitem{webb99}
Webb~N.A., Naylor~T., Ioannou~Z., Worraker~W.J., Stull~J.,
Allan~A., Fried~R., James~N.D., Strange~D. 1999, A Spatially
Resolved `Inside-out' Outburst of IP Pegasi, {\it Monthly
Notices Roy.  Astron. Soc.}, {\bf 310}, 407

\bibitem{warner95}
Warner~B. 1995, Cataclysmic Variable Stars. Cambridge:
Cambridge University Press

\end{thebibliography}
\end{document}